\documentclass{aa}
\usepackage{graphicx}
\usepackage{natbib}
\begin{document}

\title{H$_2$ active jets in the near IR as a probe of protostellar evolution
\thanks{Preprint: $http://www.mporzio.astro.it/$$\sim$$bruni/publ.html$}}

\author{A. Caratti o Garatti
    \inst{1,} \inst{2},
          T. Giannini
          \inst{1},
          B. Nisini
          \inst{1}
          \and
      D. Lorenzetti
          \inst{1}
           }

\offprints{A. Caratti o Garatti}

\institute{INAF - Osservatorio Astronomico di Roma, via Frascati 33, I-00040 Monte Porzio (Italy)
\and Universit\`a degli Studi di Roma "Tor Vergata" - Dipartimento di Fisica, via della Ricerca Scientifica 1, I-00133 Roma (Italy)
\email{caratti,giannini,nisini,dloren@mporzio.astro.it}}

\date{Received;Accepted}
\abstract{We present an in-depth near-IR analysis of a sample 
of H$_2$ outflows from young embedded sources to compare 
the physical properties and cooling mechanisms of
the different flows. The sample comprises 
23 outflows driven by Class 0 and I sources having low-intermediate luminosity.
We have obtained narrow band images in H$_2$~2.12\,$\mu$m and 
[\ion{Fe}{ii}]~1.64\,$\mu$m and spectroscopic observations in the range 1-2.5\,$\mu$m.
From [\ion{Fe}{ii}] images we detected spots of ionized gas in $\sim$74\% of the outflows
which in some cases indicate the presence of embedded HH-like objects. 
H$_2$ line ratios have been used to estimate the 
visual extinction and average temperature of the molecular gas.  
$A_{\rm v}$ values range from $\sim$2 to $\sim$15 mag; 
average temperatures range between $\sim$2000 and $\sim$4000\,K. In
several knots, however, a stratification of temperatures is found with
maximum values up to 5000\,K. Such a stratification is more commonly
observed in those knots which also show [\ion{Fe}{ii}] emission, while a 
thermalized gas at a single temperature is generally found in knots
emitting only in molecular lines. Combining narrow band imaging (H$_2$, 2.12\,$\mu$m
and [\ion{Fe}{ii}], 1.64\,$\mu$m) with the parameters derived from the spectroscopic analysis,
we are able to measure the total luminosity of the H$_2$ and [\ion{Fe}{ii}] shocked regions
($L_{H_2}$ and $L_{[\ion{Fe}{ii}]}$) in each flow. H$_2$ is the major NIR coolant with an average 
$L_{H_2}$/$L_{[\ion{Fe}{ii}]}$ ratio of $\sim$10$^{2}$.
We find that $\sim$83\% of the sources have a $L_{H_2}$/$L_{bol}$
ratio $\sim$0.04, irrespective of the Class of the driving source,
while a smaller group of sources (mostly Class I) have $L_{H_2}$/$L_{bol}$ 
an order of magnitude smaller. Such a separation reveals the non-homogeneous
behaviour of Class I, where sources with very different 
outflow activity can be found. This is consistent with other studies 
showing that among Class I one can find objects with different accretion 
properties, and it demonstrates that the H$_2$ power in the jet can be
a powerful tool to identify the most active sources among the objects
of this class.

 }

\authorrunning{A.Caratti o Garatti et al.}
\titlerunning{H$_2$ active protostellar jets}
\maketitle

%

\section{Introduction}

Matter flows emitted from young stellar objects (YSOs) are
manifestations commonly observed during all the phases of 
Pre-Main Sequence evolution: from the early accretion phase (Class
0), which lasts a relatively short time ($\sim10^4$~yrs, for a low
mass YSO), to the final contraction toward the Main Sequence
(Class II and III, $\sim10^7$~yrs). 
According to low mass star formation models, mass accretion and ejection 
rates ($\dot{M_{acc}}$,
$\dot{M_{out}}$) are expected to be strictly related, because a
significant fraction of the infalling material is ejected by the
accretion disk. The efficiency of such a coupling, however,  has to
necessarily decrease while the evolution proceeds.
Therefore the outflow/jet properties are expected to remarkably change  
with time and their study
offers both direct and indirect clues to understand the
processes related to protostellar evolution. 
In addition, several issues can be addressed by systematic observations
of jets since: ({\it i}) they signal the presence of strongly embedded driving
sources, often invisible up to far IR wavelengths; ({\it ii}) their
shape may help to reconstruct the star/disk rotation; ({\it iii})
they have dynamical and chemical effects on the closeby
interstellar material; ({\it iv}) at the same time, they are also
influenced by the properties of such material. 

Since bipolar and collimated jets are easily observed from the ground
over a wide range of frequencies, large imaging and spectroscopical
data-bases have been accumulated by various groups. Our group, in
particular, has gathered, over the last five years, a
large spectroscopical data set in the near IR (1--2.5
$\mu$m), complemented with a considerable amount of high-resolution imaging
data ([\ion{Fe}{ii}], H$_2$), on a sample of active H$_2$ outflows from
both Class 0 and Class I sources. The material collected so far has
allowed us to clarify specific aspects of the protostellar jets
physics, such as the identification of the crucial spectral range
for the study of the H$_2$ excitation conditions \citep{paper3}; the role of
[\ion{Fe}{ii}] as a diagnostic tool of embedded atomic jets \citep{paper4}; the
observational tests of state-of-art shock models \citep{paper5,mcoey,paper8}. 
Moreover, near IR spectroscopy allowed us
to investigate the properties of H$_2$ jets in individual star-forming regions
\citep[IC1396;][]{nisini1}, \citep[Vela Molecular Ridge;][]{paper6,paper1,paper7,loren}. 
Here we use our data-base to derive more general properties on the
IR activity of jets from young stars. This has been done complementing our published 
data with new observations focused mainly on the youngest
Class 0 sources.

The main aim of this study is to reveal any systematic difference
in the derived physical parameters which can be attributed to
an effect of the evolution in both the intrinsic jet properties
and in the way the jet interacts with the ambient medium.
In addition, we want to define if any relationship exists 
between the total IR cooling of the outflows 
 and the evolutionary status of the driving source.
Indeed, several authors \citep[see e.\,g.][]{cabrit0,bontemps,saraceno,andre,paper2} have
observationally shown that during the first stages of 
protostellar evolution  a correlation exists between different
tracers of the outflow activity and the source bolometric
luminosity, which is believed to be largely dominated by the
accretion luminosity ($L_{bol} \approx L_{acc} = GM_{*}
\dot{M_{in}} / R_{*}$).

In the near IR the jet shocked excited regions mainly cool through
H$_2$ quadrupole transitions, therefore the bright ro-vibrational
lines in that range represent a suitable shock tracer which can be
used to evaluate the molecular hydrogen luminosity ($L_{H_2}$) of
the outflow. Due to its very rapid cooling, H$_2$ is more
suitable to probe shock excitation and gas cooling than the CO
lines which can only give a time-integrated response \citep[see e.\,g.][]{smith1}.
The empirical classification of early
protostellar evolution by means of the H$_2$ luminosity of the
emitted jets has recently attracted much attention in different works
\citep{stanke,smith1,froebrich1,oco}, where
the simplified assumption was adopted that the total H$_2$ luminosity is 
derivable from the 2.122\,$\mu$m (1-0S(1)) line luminosity. In the
following, the validity of such an assumption as well as the most
critical aspects in deriving a reliable value of $L_{H_2}$ will be
also reviewed.

The structure of the paper is the following: in Sect.2 we define
the investigated sample of jets and associated YSOs with their
parameters; in Sect.3 our NIR observations are presented; in
Sect.4 we derive, for each jet, the physical quantities from the
molecular and ionic components; in Sect.5 a discussion is
presented in terms of jet properties vs. luminosity of the central
object. Our conclusions are summarized in Sect.6. Finally, in
Sect.7 (Appendix) the detailed results on any individual star
forming region are reported.

\begin{table*}
\caption[]{ The investigated sample
    \label{sample:tab}}
\begin{center}
\begin{tabular}{ccccccccccccc}
\hline \hline\\[-5pt]
Id & Source &  Associated &      \multicolumn{3}{c}{$\alpha$(2000.0)} & \multicolumn{3}{c}{$\delta$(2000.0)} &  Class & $L_{bol}$ & Ref.  & D   \\
   &        &   HH        &   ($^{h}$ & $^{m}$ & $^{s}$) & ($\degr$ & $\arcmin$ & $\arcsec$)        &   & ($L_{\sun}$) & & (pc) \\
\hline\\[-5pt]

1 & L1448-MM       	     & -	    	 &03&25&38.8&30&44&05.0  & 0   & 8.3--9 & 1,2 & 300  \\
2 & NGC1333-I4A    	     & -	    	 &03&29&10.5&31&13&30.5  & 0   & 14--18 & 2,1 & 350  \\
3 & HH211-MM       	     & HH211 *	    	 &03&43&56.8&32&00&50.0  & 0   & 3.6--5 & 1,2 & 300   \\
4 & IRAS05173-0555 	     & HH240/1 *  	     &05&19&48.9&-05&52&05.0 & 0/I & 17--26.6& 3,4 & 460    \\
5 & HH43-MMS       	     & HH43, HH38, HH64      &05&37&57.5&-07&07&00.0 & 0   & 3.6--5  & 3,4 & 450	     \\
6 & HH212-MM       	     & HH212 *		     &05&43&51.5&-01&02&52.0 & 0   & 7.7--14 & 1,2  & 400  \\
7 & HH26IR         	     & HH26 *		     &05&46&03.9&-00&14&52.0 & I   & 28.8   & 5 & 450	 \\
8 & HH25-MMS       	     & HH25 *		     &05&46&07.8&-00&13&41.0 & 0   & 6--7.2 & 6,1 & 450  \\
9 & HH24-MMS       	     & HH24 *		     &05&46&08.3&-00&10&42.0 & 0 &  5 & 7 & 450 	  \\
10 & IRAS05491+0247 (VLA2)   & HH111, HH311, HH113 *   &05&51&46.1&02&48&30.6 & I & 24--42 & 1,4,8 & 450  		 \\
11 & NGC2264G-VLA2  	     & - *		     &06&41&11.0&09&55&59.2  & 0 & 12--13 & 9,1  & 800    \\
12 & IRAS07180-2356 	     & HH72 *		     &07&20&10.3&-24&02&24.0 & I & 170 & 3 & 1500	 \\
13 & IRAS08076-3556 	     & HH120		     &08&09&32.8&-36&05&00.0 & I &  13--19 & 3,15 & 450    \\
14 & IRAS08211-4158 (IRS8-2) & HH219 *		     &08&22&52.1&-42&07&55.0 & I & 642& 10  & 400	  \\
15 & \#40-3 (IRS17) 	     & - *		     &08&46&32.6&-43&54&38.9  & I   & 11--245 & 11 & 700  \\
16 & BHR71-MM (IRS1)	     & HH321		     &12&01&44.0&-65&09&00.1 & 0   & 7.9--10& 2,18 & 200    \\ 
17 & BHR71 (IRS2)   	     & HH320		     &12&01&34.0&-65&08&44.0 & I  & 1--3 & 2,18 & 200	  \\
18 & IRAS12515-7641 	     & HH54		     &12&55&00.2&-76&57&00.0 & I & 0.22--0.44 & 16 & 180  \\
19 & VLA1623-243    	     & HH313 *		     &16&26&26.5&-24&24&31.0 & 0 & 1 & 2   & 160	   \\
20 & IRAS18273+0113  	     & HH460		     &18&29&49.8&01&15&20.8  & 0/I & 45--72 & 1,12  & 310  \\
21 & R CrA-IRS7     	     & HH99		     &19&01&55.3&-36&56&21.9  & I & 3.4 & 17,19 & 130	  \\
22 & L1157-MM 		     & - *		     &20&39&05.7&68&02&16.0  & 0   & 8.4--11& 1,2 & 440   \\
23 & IRAS21391+5802 	     & HH593 *		     &21&40&42.4&58&16&09.7  & 0   &150--350 & 13,14 & 750  \\
\hline \hline
\end{tabular}
\end{center}
{\bf References}: {\bf (1)} \citet{froeb}, {\bf (2)} \citet{andre}, {\bf (3)} \citet{reipurth}, {\bf (4)} \citet{molinari},
{\bf (5)} \citet{davis1}, {\bf (6)} \citet{lis}, {\bf (7)} \citet{chini1}, {\bf (8)} \citet{chini2}, {\bf (9)} \citet{ward},
{\bf (10)} \citet{paper6}, {\bf (11)} \citet{paper7}, {\bf (12)} \citet{larsson}, {\bf (13)} \citet{beltran},
{\bf (14)} \citet{nisini1}, {\bf (15)} \citet{persi}, {\bf (16)} \citet{Cohen}, {\bf (17)} \citet{marraco},
{\bf (18)} \citet{bourke}, {\bf (19)} \citet{wilking0} \\
Note: asterisks in column 3 indicate outlows with a defined morphology, where the knot assignation taken
from literature or derived from this work (see Appendix) is certain.\\
\end{table*}


\section{The investigated sample}
\label{sample:sec}

The investigated sample of H$_2$ jets is presented in
Table~\ref{sample:tab}. The driving selection criterion is the
occurrence of jet activity in the near IR: for each selected
source an IR jet has been detected in previous H$_2$ 2.12 $\mu$m
narrow band imaging at a sensitivity level adequate to perform
low resolution near IR spectroscopy with currently available
instrumentation. The second criterion is the existence of a 
recognized exciting source whose bolometric luminosity is well
ascertained and typical of low to intermediate mass young objects
($0.2 L_{\sun} \le L_{bol} \le 640 L_{\sun}$). An important
consequence of our criteria is that the selected targets are
associated only with Class 0 and I driving sources, while objects
belonging to more evolved classes remain unselected.
Thus the jets exhibiting strong enough H$_2$ emission are those
related to deeply embedded protostars. In particular, our sample is
constituted by 23 H$_2$ active jets, whose driving sources are
distributed as 12 Class 0, 9 Class I and 2 intermediate Class
(0/I), according to the literature \citep[see e.\,g.][]{froeb}.
Table~\ref{sample:tab} lists the exciting source and its correlated HH,
the coordinates of the exciting source, its
evolutionary Class, bolometric luminosity and distance. The given
range of $L_{bol}$ for each source stems from different
determinations given in the literature. All the targets belong to
nearby star forming regions ($\le$ 800\,pc), with the exception of
$IRAS07180$--$2356$ located at 1500\,pc.

Not all the outflows of the sample show a clear bipolar structure, being sometimes located
in crowded regions where jets from different YSOs are present. In Table~\ref{sample:tab} column 3,
an asterisk indicates those outflows (14 out of 23) that exhibit a clear morphology and all the observed
H$_2$ knots can be associated with the same driving source (see also Sec.~4.1,~4.3 and the Appendix).

\begin{table*}
\caption[]{ Journal of observations - Imaging and Spectroscopy
    \label{dati:tab}}
\begin{center}
\begin{tabular}{ccccccc}
\hline \hline\\[-5pt]
Object   &  Telescope/  & Date of obs. & Imaging  & \multicolumn{3}{c}{Spectroscopy} \\
         &  Instrument  & (d,m,y) &  Band   & Wavelength($\mu$m)& $\mathcal R$ & P.A.($\degr$)  \\
 \hline\\[-5pt]
L1448     &  TNG / NICS &  16/10/2003 & [\ion{Fe}{ii}], H$_{c}$ &  & & \\
NGC1333    & TNG / NICS &  17/10/2003 & [\ion{Fe}{ii}], H$_{c}$, H$_2$, K$_{c}$ & 1.15--2.50 & 500 & 34 \\
HH211     & TNG / NICS &  17/10/2003 &[\ion{Fe}{ii}], H$_{c}$, H$_2$, K$_{c}$ & &  & \\
       & AZT-24/SWIRCAM & 23/10/2004 & & 1.45--2.30 & 200 & 90 \\
HH240/1     & AZT-24/SWIRCAM & 17/11/2004 & [\ion{Fe}{ii}], H & & & \\
HH212       &  ESO-NTT / SofI & 11/01/2001 & H$_2$, K$^{\prime}$ & 0.95--2.50 & 600 & 22,24 \\
        & AZT-24/SWIRCAM & 16/11/2004 & [\ion{Fe}{ii}], H & & & \\
HH26        & AZT-24/SWIRCAM & 17/11/2004 & [\ion{Fe}{ii}], H & & & \\
HH25        & AZT-24/SWIRCAM & 17/11/2004 & [\ion{Fe}{ii}], H & & & \\
HH24        & AZT-24/SWIRCAM & 17/11/2004 & [\ion{Fe}{ii}], H & & & \\
HH111       & ESO-NTT / SofI & 11/01/2001 & H$_2$, K$^{\prime}$ & & & \\
NGC2264G    &  ESO-NTT / SofI & 13/03/2003 & H$_2$, K$^{\prime}$ & 0.95--2.50 & 600 & 84 \\
        & AZT-24/SWIRCAM & 17/11/2004 & [\ion{Fe}{ii}], H & & & \\
BHR71       & ESO-NTT / SofI & 02/05/2001 & [\ion{Fe}{ii}], H$_2$, K$^{\prime}$ & & & \\
        & ESO-VLT / ISAAC &  15/07/2002 &  H$_2$, K$^{\prime}$   & & & \\
HH54        & ESO-NTT / SofI  &  05/06/1999 &  H$_2$, K$^{\prime}$   & & & \\
        & ESO-VLT / ISAAC &  01/01/2005 &  H$_2$, K$^{\prime}$   & & & \\
VLA1623--HH313    & ESO-NTT / SofI & 11/01/2001 & H$_2$, K$^{\prime}$ & 0.95--2.50 & 600 & 298,299,305 \\
IRAS18273+0113  & ESO-NTT / SofI & 05/06/1999 & H$_2$, K$^{\prime}$ & 1.50--2.50 & 600 & 328 \\
        & ESO-VLT / ISAAC & 02/07/2001 & [\ion{Fe}{ii}], H, H$_2$, K$^{\prime}$ &  &  & \\
HH99         & ESO-NTT / SofI & 05/06/1999 & H$_2$, K$^{\prime}$ & & & \\
         & ESO-VLT / ISAAC & 15/06/2005 & H$_2$, K$^{\prime}$ & & & \\
L1157          &  TNG / NICS &  16/10/2003 & [\ion{Fe}{ii}], H$_{c}$ & 1.15--2.50 & 500 & 158,356 \\
IC1396N    &  TNG / NICS &  17/10/2003 & [\ion{Fe}{ii}], H$_{c}$, H$_2$, K$_{c}$ & 1.45--2.50 & 500 & 79 \\
\hline \hline
\end{tabular}
\end{center}
\end{table*}

\section{Observations and data reduction}

\subsection{Imaging}

The observations presented in this paper were collected during several runs between 1999 and 2005, at four different
telescopes, namely ESO-VLT and ESO-NTT (Chile), Telescopio Nazionale Galileo (TNG) (Canary Islands-Spain) and AZT-24
(Campo Imperatore-Italy) (see Table~\ref{dati:tab}).
In addition, data from the ESO science archive facility\footnote{http://archive.eso.org/} have been retrieved for
HH24, BHR71 and IRAS18273+0113.

We used narrow band filters centered on the H$_2$ (2.12\,$\mu$m)
and [\ion{Fe}{ii}] (1.64\,$\mu$m) lines to detect both molecular
and ionic emission along the outflows. Additional broad band ($H$
and $K$) or narrow band imaging at nearby wavelength positions
have been gathered to remove the continuum.
Table~\ref{instruments:tab} summarizes the characteristics of the
different near IR cameras used for our observations.

All the raw data were reduced by using IRAF\footnote{IRAF (Image Reduction and Analysis Facility) is
distributed by the National Optical Astronomy Observatories, which are operated by AURA, Inc., cooperative agreement with the
National Science Foundation.} packages applying standard procedures for sky subtraction, dome
flat-fielding, bad pixel and cosmic rays removal and imaging mosaic.

Calibration was obtained by means of photometric standard stars observed in both narrow and broad band filters.

Once the narrow band images were calibrated and continuum-subtracted, we detected and measured H$_2$ and
[\ion{Fe}{ii}] fluxes on each knot using the task {\em polyphot} in IRAF, defining each region within a
$2 \sigma$ contour level above the sky background.

\subsection{Spectroscopy}

Low resolution spectroscopy was acquired during four different runs
at three telescopes, namely ESO-NTT, TNG and AZT-24. As reported in
Table~\ref{dati:tab}, we obtained long slit spectra for eight Class 0 jets, with different position angles (P.A.),
mostly oriented along each jet axis (except for HH211, where the slit encompasses only the bow-shock in the blue lobe).
The spectra of the remaining jets of our sample (with the exception of L1448, where we have no spectra) have been
published in our previous papers.

The covered spectral range and resolution are also indicated in Table~\ref{dati:tab}.
To perform our spectroscopic
measurements, we adopted the usual ABB`A' configuration, for a
total integration time between 1200~s and 3600~s per slit. Each
observation was flat fielded, sky subtracted and corrected for the
curvature derived by long slit spectroscopy, while atmospheric
features were removed by dividing each spectrum by a telluric
standard star (O spectral type), normalized to the blackbody
function at the star's temperature and corrected for hydrogen
recombination absorption features intrinsic to the star. The
wavelength calibration was retrieved from Xenon and Argon lamps.
The flux calibration was obtained from the observation of
different photometric standard stars of the catalogues by
\citet{carter} and \citet{persson}.

The obtained line fluxes were compared with those available in the literature: HH211
\citep[][]{ocon}, HH212 \citep[][]{zinn,tedds}, HH313 \citep[][]{eis}.
 The results are consistent with our values considering the calibration errors and the different slit widths.

\begin{table}
\caption[]{ Camera characteristics
    \label{instruments:tab}}
\begin{center}
\begin{tabular}{cccc}
\hline \hline\\[-5pt]
Instrument   &  FoV  & pixel scale & Ref. \\
             &  ($\arcmin \times \arcmin$) & ($\arcsec$/pix) &  \\
\hline\\[-5pt]
ISAAC    & 2.5 $\times$ 2.5 & 0.148 & 1 \\
SofI     & 4.9 $\times$ 4.9 & 0.290 & 2 \\
NICS     & 4.2 $\times$ 4.2 & 0.250 & 3 \\
SWIRCAM  & 4 $\times$ 4     & 1     & 4 \\

\hline \hline
\end{tabular}
\end{center}
{\bf References}: {\bf (1)} \citet{cuby}, {\bf (2)} \citet{lidman}, {\bf (3)} \citet{baffa}, {\bf (4)} \citet{dalessio}\\
\end{table}


\section{Results \& data analysis}
\subsection{[\ion{Fe}{ii}] and H$_2$ imaging}

In this section, the results and the analysis obtained from narrow band images are presented.
The 1.64\,$\mu$m images were obtained to discover the presence of 
ionised jets or dissociative
regions in bow-shocks associated with known molecular outflows and to compute the [\ion{Fe}{ii}] 
radiative contribution
to the overall energy budget.
The [\ion{Fe}{ii}] emission regions are always localized in  compact spots, representing only a small 
fraction of the region emitting in H$_2$.
On the new [\ion{Fe}{ii}] images of the 13 jets presented in this
paper (see Table~\ref{dati:tab}), we observe iron emission in 9
objects, namely L1448, HH211, HH240-241, HH212, HH24, HH26,
NGC2264G, L1157 and IC1396N. Moreover, considering the entire
investigated sample, of 23 outflows, we detect [\ion{Fe}{ii}]
emission in 17 objects (74\%), 9 out of 12 from Class 0 (75\%), 1
out of 2 from Class 0/I (50\%) and 7 out of 9 from Class I YSOs
(78\%). This indicates that the iron emission is homogeneously
distributed over jets of the different YSOs classes and that the
presence or not of [\ion{Fe}{ii}] in the jet is not strictly
correlated with the evolutionary stage of the exciting source.

In our analysis of the sample, H$_2$ images were mainly acquired for photometric purposes.
In a few cases, however, the images were also utilized in conjunction with images of the 
same region taken at
different epochs, to derive the proper motion of the knots (see VLA1623, HH54 and HH99 in the Appendix).
The kinematical information was particularly useful in those cases where several jets are 
present in the same region.
In such cases it has been possible to disentangle which knots are associated with the source 
under consideration.
The newly discovered knots (both molecular and ionic), along with their positions and fluxes, are listed
in Table~\ref{knots:tab}\footnote{Tables~\ref{knots:tab}-\ref{IC1396sp:tab} are only available in
electronic form at http://www.edpsciences.org}.

In the Appendix, together with a short description,
we report the images of those objects not yet published in the literature (mainly [\ion{Fe}{ii}] 
emission), or that give some additional information and details on the morphology of the outflows.

\begin{figure*}
 \centering
   \includegraphics [width=12 cm] {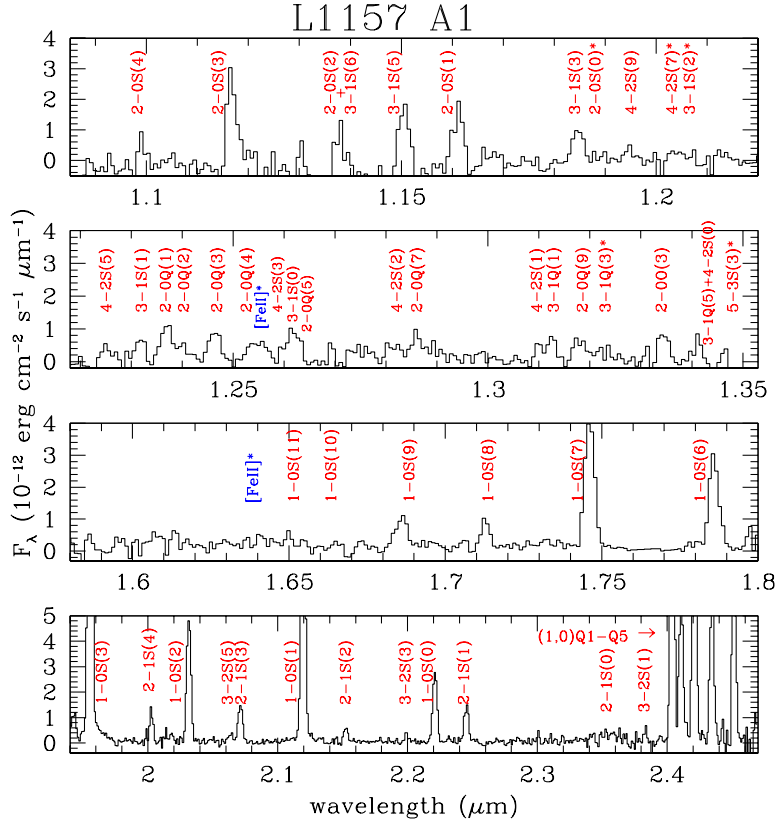}
   \caption{ 1.1--2.5\,$\mu$m low resolution spectrum of knot A1 in the L1157 outflow.
   An asterisk near the line identification marks the detections between 2 and 3 sigma.
\label{L1157_A1_sp:fig}}
\end{figure*}

\begin{figure*}
 \centering
 \includegraphics [width=12 cm]{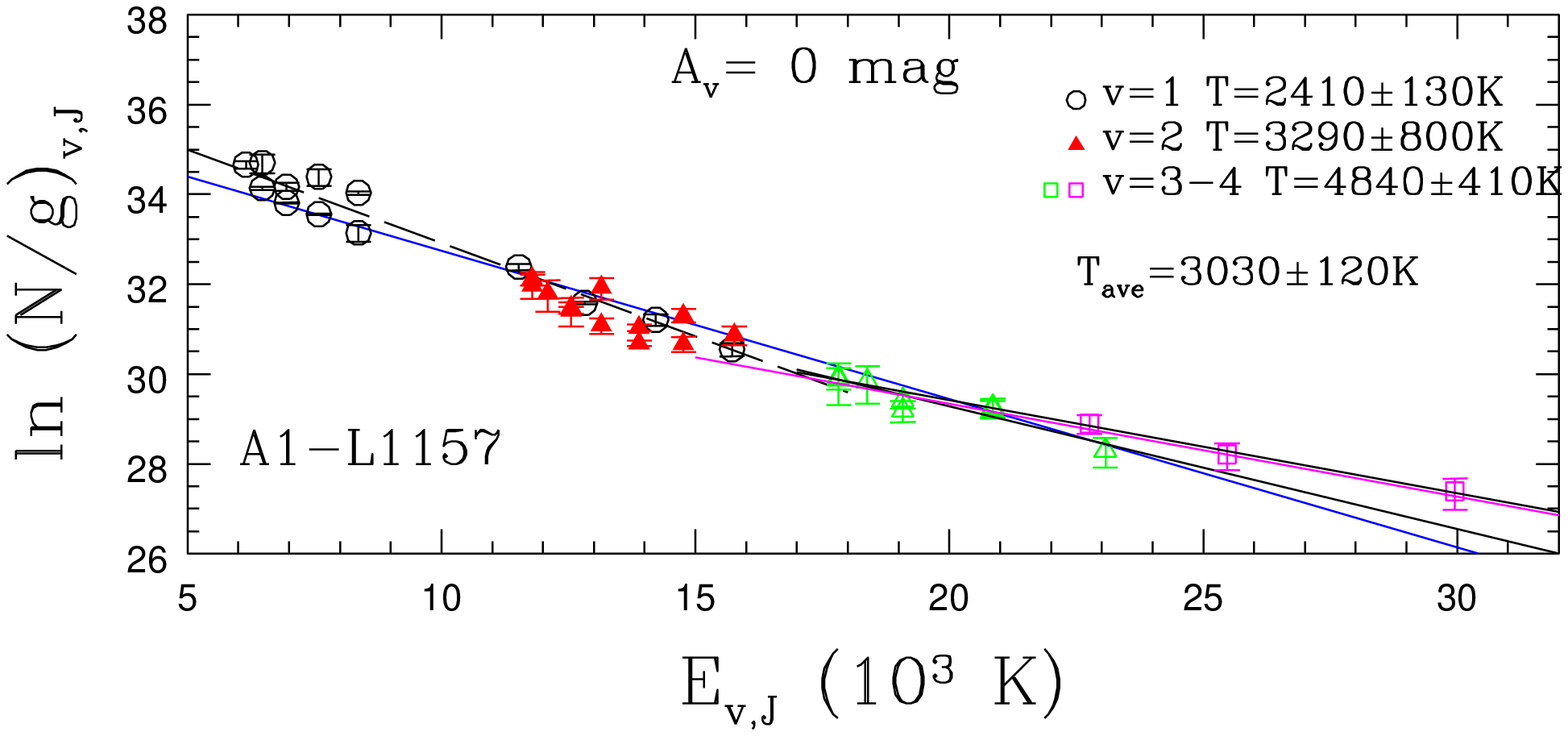}
  \caption{ Rotational diagrams of knot A1 in the L1157 outflow derived from low resolution spectroscopy.
   Different symbols indicate lines coming from different vibrational levels, as coded in the upper right corner of
   the box. The straight lines represent the best fit through the v=1, v=2,3,4 (solid line). 
   The derived temperature is also indicated in the upper right corner of the box.
\label{rotL1157:fig}}
\end{figure*}


\subsection{Spectroscopy}

In Fig.~\ref{L1157_A1_sp:fig} we show the spectrum
of knot A1 in L1157, which can be considered as representative of the whole sample.
The most prominent features detected are the H$_2$ lines coming from different
upper vibrational levels, ranging from $v=1$ to $v=3$. In some cases (HH313 A in VLA1623 and some knots in L1157) we also detect higher
vibrational lines ($v=4$ and 5) coming from energy levels up to 30\,000\,K.

[\ion{Fe}{ii}] emission is detected in 20\% of the observed knots, that is $\sim$63\% of the objects (5 of 8 spectra reported in
Table~\ref{dati:tab}, namely HH211, HH212, VLA1623, IRAS20386, IRAS21391),
showing the transitions at 1.644\,$\mu$m and (often) at 1.257\,$\mu$m, while the faintest lines (1.534, 1.600, 1.678\,$\mu$m)
are detected just in a few knots (NK1 and SK1 in HH212, HH313 A in VLA1623).
Moreover, in HH313 A, we observe (not resolved) the [\ion{C}{i}] doublet at 0.983 and 0.985\,$\mu$m.
No other ionic feature is revealed in the analysed spectra, mostly due both to the extinction ($A_{\rm v}$$\ge$5~mag)
and to the low excitation of the observed knots.

In Tables (\ref{NGC1333sp:tab}-\ref{IC1396sp:tab}) we list the fluxes (uncorrected for the extinction)
of the identified lines together with their vacuum wavelength. Line fluxes have been obtained by fitting the profile with
a single or double Gaussian in case of blending. The uncertainties associated with these data are derived only
from the rms of the local baseline multiplied by the linewidths (always comparable with the instrumental profile width). Lines
showing fluxes with a S/N ratio between 2 and 3, as well as those blended, have been labelled. Additional uncertainties
in the fluxes are derived from the absolute calibration ($\approx$ 10\%).

\subsubsection{Deriving $A_{\rm v}$ and H$_2$ temperature}

The main quantitative information we can gather from the observed
line emission is the reddening towards each knot and the gas
temperature characterizing the H$_2$ emission. In the few cases
where [\ion{Fe}{ii}] at 1.53 $\mu$m or 1.60 $\mu$m has been observed
in addition to the 1.64 $\mu$m line, the electronic density
$n_{\rm e}$ also can be derived \citep{paper3}.

We have employed all the available H$_2$ ratios to simultaneously 
evaluate extinction and gas temperature in the
framework of ro-vibrational diagrams, where the extinction-corrected 
column densities, divided by their statistical weights, are plotted 
against their excitation energies. If the gas is thermalized, 
the points in the diagram align in a straight line, whose angular
coefficient is the reciprocal of the gas temperature.
Line pairs originating from the same level should lie in the same position
in the diagram: by varying the $A_{\rm v}$ value and increasing the 
goodness of the fit (maximising the correlation coefficient) 
extinction and temperature can be simultaneously evaluated \citep[see e.\,g.][]{paper4,davis4}.
To minimize the uncertainties, only the transitions with a S/N$>$3 
and not affected by blending with other lines have been used.
If the extinction is relatively low ($A_{\rm v}$$\leq$5~mag), several bright pairs (e.\,g.
2-0S(i)/2-0Q(i+2) and 3-1S(i)/3-1Q(i+2) with i=0,1,2...) lying
between 1 and 1.35\,$\mu$m ($J$ band) are observed. In these cases  
we are able to reduce the uncertainty on $A_{\rm v}$ to about 1\,mag.
On the contrary, where the extinction is high ($\ge$ 10\,mag), lines in the $J$
band are not visible and the only available pairs involve lines in the range 2.4 -2.5 $\mu$m,
which are strongly affected by poor atmospheric transmission. In these cases the uncertainty
on $A_{\rm v}$ is 5-10\,mag.
In Fig.~\ref{rotL1157:fig}, as an example, we show the ro-vibrational 
diagram of knot A1 in L1157 where the lines coming from different vibrational 
levels are indicated with different symbols.
In this diagram a single temperature LTE gas fits all the lines fairly well. In some cases, 
however, there is evidence of a stratification in the gas temperature, in which case
different temperatures are derived by fitting the various  
vibrational states. In such cases, a single fit through all the lines
gives only a measure of the `averaged' temperature.

In columns 3 and 4 of Table~\ref{phys:tab} we report the range of average temperatures 
and extinctions derived in {\it each} jet of the
sample (see Table~\ref{sample:tab}), while the values for the knots studied in this paper for the first time
are given in Table~\ref{phys_knots:tab}. 
Average temperatures range from 1800 to 4200\,K, with typical values around 2500--3000\,K, while the extinction
spreads from 0 to 15\,mag, with the exception of the IRS17 jet where a value of 30\,mag is found near the exciting source.

In those cases where the [\ion{Fe}{ii}] emission is observed, the
ratio between the 1.257 and 1.644\,$\mu$m bright lines has been also
used to independently evaluate the extinction. We obtain, for 14 knots
of our sample, $A_{\rm v}$  values that on average are 2.5 mag higher than
those derived from the H$_2$ lines. Such result, however, must to
be taken with some caution, since the 1.257/1.644 ratio 
is suspected to systematically overestimate the extinction value \citep{nisini05}.

\subsection{H$_2$ and [\ion{Fe}{ii}] luminosities}
As a consequence of the high level of emission of H$_2$ and [\ion{Fe}{ii}] lines in our NIR spectra, 
we expect that the luminosities of these two species ($L_{H_2}$ and $L_{[\ion{Fe}{ii}]}$)
represent a significant fraction of the overall energy radiated away during the gas cooling
\citep[between 10\% and 50\%, see e.\,g.][]{paper1}, while most of the remaining
energy is emitted at MIR and FIR wavelengths \citep[see e.\,g.][]{paper1,paper2}.
A powerful and commonly used way to estimate $L_{H_2}$ is to directly derive it from
the 2.12\,$\mu$m imaging, by considering $L_{2.12}$ as a fraction of $L_{H_2}$, whose specific value
depends on the gas temperature (in particular $L_{H_2} \sim 0.1 \times L_{H_2}$, if $T \sim 2000$\,K).
This approximation suffers, however, from two main limitations: {\it i}) the
lack of knowledge of the extinction, which affects the 2.12\,$\mu$m luminosity;
{\it ii}) the presence of different temperature components, as often seen in the 
rotational diagrams.

The $L_{H_2}$ uncertainty due to the extinction can be quantitatively evaluated
by considering that, in our sample, $A_{\rm v}$ ranges typically from 0 to 15 mag (considering 
also the uncertainties on $A_{\rm v}$, this range is enlarged to 0-25 mag). 
This implies that, if no extinction is applied, $L_{2.12}$ could be underestimated by up to an order of magnitude.
If, however, as in our case, the extinction is measured in spectroscopic
observations, the uncertainty on the 2.12\,$\mu$m intrinsic luminosity is significantly reduced, 
from 10\% to 300\%, if the uncertainty on $A_{\rm v}$ is 1 and 10 mag, respectively.

\begin{figure}
 \centering
 \includegraphics [width=6 cm]{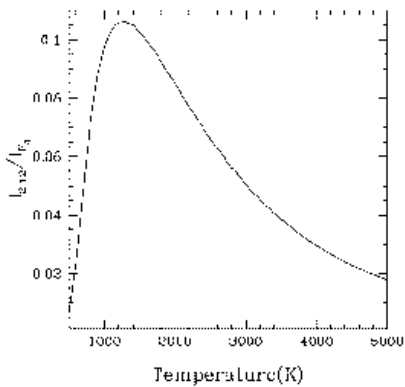}
  \caption{ The plot represents the intensity of the 2.12\,$\mu$m line with respect to the H$_2$ total intensity
  as a function of the gas temperature in LTE conditions.
\label{212vsT:fig}}
\end{figure}

To evaluate the uncertainty on $L_{H_2}$ introduced by the presence of different temperature components,
we have plotted in Fig.~\ref{212vsT:fig} the 2.12\,$\mu$m line intensity as a function of the gas temperature
in LTE conditions: here it can be noticed how the approximation $L_{2.12} \propto 0.1 \times L_{H2}$ is adequate
in a range of temperatures from 1500 to 2500 K. In some of
the investigated knots, however, average temperatures higher than 3000\,K have been measured (see Table~\ref{phys:tab}),
in which cases $L_{2.12}$ represents less than 4\% of the total $L_{H_2}$.

The above considerations have lead us to obtain an accurate measurement of $L_{H_2}$. We have applied 
the following procedure: {\it i}) we associate each emission knot, traced by narrow 
band imaging (obtained by us or available in the literature), with the driving YSO, by means of the morphological 
structure and/or kinematical data (3 objects, see Appendix). Alternatively, when the assignation
of a knot remains uncertain, a larger error on $L_{H_2}$ has been applied\footnote{ As an example, some knots in BHR71,
namely HH321 A and knot~6 \citep[see Fig.~3 in][]{paper5}, have been assigned to both sources (IRS\,1 \& 2) increasing
the uncertainty on $L_{H_2}$ by the knot luminosities ($\sim$10$^{-2}$\,L$_{\sun}$).}; 
{\it ii}) the flux of each knot has been evaluated by 
photometry from the 2.12 narrow band image; {\it iii}) once having derived the physical parameters 
($A_{\rm v}$, $T$) of single knots from NIR spectroscopy, we dereddened the 2.12\,$\mu$m flux,
adopting the \citet{riek} reddending law, and computed the line ratios with the other H$_2$ lines
by applying a simple radiative LTE code. From such ratios, the absolute intensities of individuals lines were 
computed and added to obtain the total H$_2$ luminosity.

\begin{figure*}
 \centering
 \includegraphics [width=12 cm]{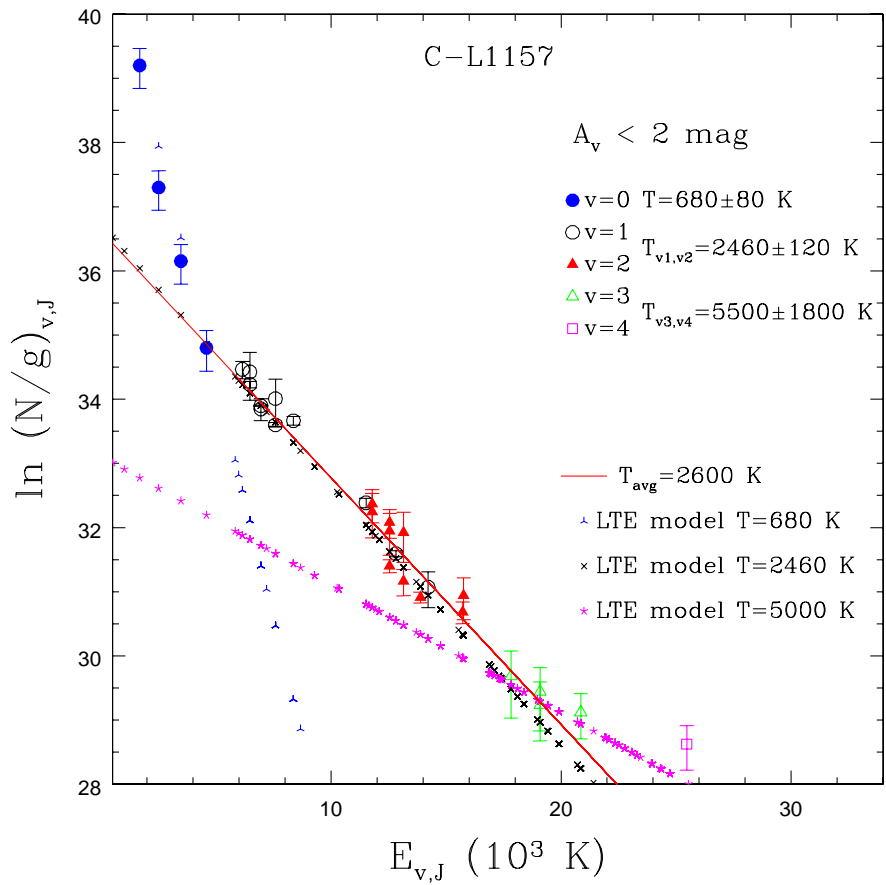}
  \caption{ Rotational diagrams of knot C in L1157. The datapoints from different vibrational levels \citep[including v=0 from][]{cabrit},
  are represented with larger symbols and have error bars. LTE H$_2$ models (at T = 680, 2460 and 5000\,K) are reported with smaller symbols.
  The straight line indicates the average temperature adopted (at T = 2600\,K).
  \label{temperature:fig}}
\end{figure*}

The adopted LTE code computes the line intensities involving levels with $0 \le v \le 14$ and $0 \le J \le 29$
(E$_{v,J}$ $\le 50\,000\,K)$. Rovibrational energies have been taken from \citet{dab} and the Einstein coefficients from \citet{wol}. 
We always assume an ortho/para ratio equal to three. To check the validity of our approximation, we compared to
the $L_{H_2}$ luminosity computed in LTE at a single temperature with that derived by considering multiple temperature components.
An example is given in Fig.~\ref{temperature:fig}, which plots the rotational diagram of the knot C in L1157 constructed
combining our near-IR data with the $v$=0 mid-IR lines observed by ISOCAM \citep{cabrit}. We can identify
three different gas components  probed by the $v$=0, $v$=1,2 and $v \ge$3 transitions and corresponding to the excitation energies 
$E_{v,J}\le 6000$\,K, $6000 \le E_{v,J} \ge 20000$\,K
and $E_{v,J} > 20000$\,K, respectively. The fitted temperature for the `cold', `warm' and `hot' components is 700, 2500 and 5000\,K,
while, by fitting through only the NIR datapoints, we obtain $T_{avg} = 2600$\,K. The contribution to $L_{H_2}$ coming from each
component has been evaluated and compared with that computed if the average temperature is assumed. While the luminosity
of lines with $E_{v,J} > 6000$\,K changes of only few percent if the fitted or the average temperature is adopted,  
the luminosity of the pure rotational lines is slightly underestimated ($\sim$10-15\%) if computed by assuming $T_{avg}$:
this can be seen in Fig.~\ref{temperature:fig} by observing that the straight line,
which represents the fit through all the NIR data, lies systematically below the points with energy less than 6000\,K. 
However, since  this systematic underestimate of $L_{H_2}$ is negligeable with respect to the uncertainties 
introduced by the $A_{\rm v}$ determination, we have not corrected our derived LTE values for it.

A complementary way to check how much our $L_{H_2}$ estimate reproduces the real cooling in the considered
regions is to compare this value with that obtained by modelling the H$_2$ observed lines with
a shock model in which the stratification of the different physical parameters is taken into account.
While it is beyond of the scope of this paper to obtain a rigourous modelling for all the
observed jets in our sample, we can make this comparison in the few flows we have been salready
analyzed through shock models in our previous papers \citep{paper5,mcoey,paper8}. We checked that in all these cases
our estimated $L_{H_2}$ luminosity is within 15-20\% of the luminosity derived from a rigorous shock model fitting and
therefore, it represents a good approximation of the $H_2$ total cooling.

As regards $L_{[\ion{Fe}{ii}]}$, we have followed the same steps as for the $L_{H_2}$ estimate, by employing a  
NLTE model to compute the line ratios with respect to the 1.64\,$\mu$m line. 
The code considers the first 16 fine structure levels of \ion{Fe}{ii} \citep[see][for details]{paper4}. 
The input parameters are the electron density ($n_{\rm e}$) and the gas temperature.
The first has been derived for some of the considered knots by \citep{paper4}, for the others we have assumed a range of 
$10^{3}-10^{5}$\,cm$^{-3}$, which is typical of the environments where the [\ion{Fe}{ii}] emission arises. Since all the observed
lines have very similar excitation energy ($\sim$ 11\,000-12\,000\,K) their ratio is not sensitive to the gas temperature;
therefore an average value of 10\,000\,K has been assumed for all the emitting regions; we find that
$L_{[\ion{Fe}{ii}]}$ changes by a factor of two in the range of temperatures between 5\,000--10\,000\,K.

In Table~\ref{phys:tab} the results of our analysis are reported.
For each jet, we give the observed range of H$_2$ temperatures and extinctions (columns 3 and 4),
along with the H$_2$ and [\ion{Fe}{ii}] luminosities (columns 5 and 6). $L_{H_2}$ is, on average, two order of
magnitudes greater than $L_{[\ion{Fe}{ii}]}$, that is, in our sample of jets the cooling occurs mostly through H$_2$ line
emission.

\begin{table*}
\caption[]{ Parameters of the jets of the sample derived from the H$_2$ and [\ion{Fe}{ii}] line analysis.
    \label{phys:tab}}
\begin{center}
\begin{tabular}{cccccc}
\hline \hline\\[-5pt]
Id & Outflow/HH & Temperature & $A_{\rm v}(H_2)$  &  $L_{H_2}$    & $L_{[\ion{Fe}{ii}]}$ \\
   &        &    (K)      & (mag)  & (10$^{-1}$ L$_{\sun}$) & (10$^{-3}$ L$_{\sun}$)  \\
\hline\\[-5pt]

1 & L1448                          &   $\cdots$   & $\cdots$ &      1.1$\pm$0.4       &   $1.5^{+0.9}_{-0.7}$         \\
2 & NGC1333                        &  2000--2800  &  5--15   &      1.0$\pm$0.3       &  not detected   \\
3 & HH211                          &  2500--2800  &  5--10   &      1.1$\pm$0.3       &  3.1$\pm$2   \\
4 & L1634 (HH240/1)                &  2000--4000  &   2--5   &      2.4$\pm$0.3       &  $39^{+7}_{-4}$ \\
5 & HH43                           &    4000      &  1--3    &      $<$6          &   $\cdots$   \\
6 & HH212                          &  1700--3000  &  3--13   &      1.2$\pm$0.4       & 3.2$\pm$0.4   \\
7 & HH26                           &  2350--3500  &  0--3    &      2.9$\pm$0.5       &  $2^{+1.8}_{-0.5}$   \\
8 & HH25                           &  2000--2800  &  0--8    & $1.0^{+0.5}_{-0.2}$    &   not detected  \\
9 & HH24                           &  2200--3500  &  0--8    & $1.4^{+1.6}_{-0.5}$    &  $2^{+1.8}_{-0.5}$   \\
10 & HH111                         &  2200--3500  &  5--11   & $0.5^{+0.2}_{-0.09}$   &  14.5$\pm$7   \\
11 & NGC2264G                      &  2100--2800  &  3--8    & $2.8^{+0.6}_{-1.3}$    &  12$\pm$6   \\
12 & HH72                          &  2100--3400  &  5--15   & $7.6^{+0.7}_{-0.6}$    &   8$\pm$2  \\
13 & HH120                         &  2200--4000  &  1--5    & $2.0^{+1.0}_{-1.1}$    &   5$\pm$1  \\
14 & IRS8-2 (HH219)                &  2000--4200  &  2--15   &    0.7$\pm$0.1         &    $22^{+6}_{-11}$ \\
15 & IRS17                         &  1900--3500  &  5--30   & $4.1^{+1.6}_{-0.9}$    &    $12^{+15}_{-7}$    \\
16 & BHR71 (HH321)                 &  2200--4000  &  0--2    & $1.2^{+0.5}_{-0.8}$    &  not detected   \\
17 & BHR71 (HH320)                 &  2200--3900  &  0--2    & $0.6^{+0.5}_{-0.4}$    &  not detected   \\
18 & HH54                          &  2500--3300  &  1--3    & 0.18$\pm$0.03          & 1.6$\pm$0.5  \\
19 & VLA1623-243 (HH313)           &  2000--3100  &  4--15   & $0.40^{+0.06}_{-0.14}$ & 0.5$\pm$0.2  \\
20 & IRAS18273+0113                &  2200--2600  &  5--10   &       0.4$\pm$0.2      & not detected     \\
21 & HH99                          &  2000--3600  & 4    &  0.07$\pm$0.04         &   6$\pm$2   \\
22 & L1157                         &  2100--3100  & 0--2     & 1.7$\pm$0.1        & 0.7$\pm$0.2  \\
23 & IC1396N                       &  2400--2700  &  5--15   & $6.1^{+3.8}_{-1.6}$    & $20^{+18}_{-7}$ \\

\hline \hline
\end{tabular}
\end{center}
Note: the ranges (min-max) of $T$ and $A_{\rm v}$ observed in the
outflow knots are reported.
\end{table*}


\section{Discussion}
\subsection{Comparing the sampled jet properties}

Our sample of H$_2$ active jets is comprised of objects belonging to different
star forming regions and excited by sources of different ages. 
This circumstance allows us to investigate whether the physical parameters
and chemical structure of the different jets depend on
the evolution of the driving source and/or on the environment. In particular, 
an evolutionary trend in the jets from a molecular to an ionic composition 
is expected mainly due to changes in the density of both the
jet and the ambient medium where the jet propagates \citep[see e.\,g.][]{smith0,smith1}.
Indeed jets from Class 0 sources travel in the high density gas 
where most young protostars are embedded. In this framework, non-dissociative C-type shocks 
that cool mainly through molecular lines, should be favoured . 
On the other hand, in older protostars
(Class I) the jet propagates in a medium at lower density,
due to the fact that previous mass loss events have already swept out
the ambient gas. In these conditions, dissociative J-type shocks
should be favoured due to the lower influence of the local magnetic fields
whose strength is a function of the density \citep[see e.\,g.][]{hollenbach}. 
Most of the outflows of our sampled Class 0 are not associated with 
optical HH, which can be evidence of the scenario described above.
However, in the embedded regions surrounding Class 0 sources,
optical HHs could be hidden by the extinction, thus remaining undetected.
Our images in [\ion{Fe}{ii}] 1.64\,$\mu$m can be used to signal the
presence of embedded dissociative shocks, thus studying if there
is a real difference in the characteristics of the shocks between
the two classes of sources. 
Indeed, we detected [\ion{Fe}{ii}] emission
spots in $\sim$74\% of the investigated sample, irrespective
of the driving source class. Moreover, the NIR spectra in our sample do
not exhibit any substantial difference, showing both molecular
(H$_2$) and ionic emission ([\ion{Fe}{ii}], [\ion{C}{i}],
[\ion{S}{ii}]) in both Class 0 and Class I flows. 
Also a more quantitative analysis of the derived jet parameters in
Table~\ref{phys:tab} does not indicate substantial differences
between the two groups. Comparing the logarithmic ratio between
H$_2$ and [\ion{Fe}{ii}] outflow luminosities
($log(L_{H2}/L_{FeII})$) in both classes of the sample, we obtain
similar values in the younger (2.1$\pm$0.6) with
respect the older jets (1.3$\pm$0.7). In addition, if we consider
the maximum temperature for both source classes, on average, Class
0 outflows have similar H$_2$ temperatures (3200$\pm$500\,K) to Class I outflows (3600$\pm$400\,K).

This analysis shows that jet physical properties, as a function of age,
cannot be determimed when comparing Class 0 outflows with outflows of `young'
Class I sources, as in our case. Differences are more
pronounced only when considering sample of sources with a larger spread
in age (e.\,g. Class 0 protostars with `evolved' Class I and T Tauri stars).

\subsection{$L_{H_2}$ and bolometric luminosity}

Several works in the past decade have discussed the interplay between the
evolutionary properties of young protostars (Class 0/I) and the
strength of their associated outflows. \citet[][]{bontemps} were
among the first to show that Class 0 protostars drive outflows more
powerful than those of Class I sources of the same luminosity, 
correlating the CO momentum flux ($F_{CO}$) of a sample of embedded protostars
with their bolometric luminosity ($L_{\rm bol}$). The larger $F_{CO}$/$L_{\rm bol}$
efficiency in Class 0 sources has been interpreted as due to the
strict relationship between mass accretion and mass loss rates, and
to the fact that the fraction of total luminosity due to accretion is
larger for Class 0 than the Class I sources. 
A similar result has been found by \citet{paper2} using
as an indicator of the outflow power the total luminosity radiated 
in the far IR by OI, H$_2$O and CO, which represent, together with H$_2$, 
the major coolants in the dense shocks occurring in outflows. 
In particular, \citet{paper2} found that $L_{\rm FIR}$/$L_{\rm bol}$
change from $\sim$10$^{-2}$ to $\sim$10$^{-3}$ evolving between Class 0
and Class I sources.

\begin{figure*}
 \centering
 \includegraphics [width=12 cm]{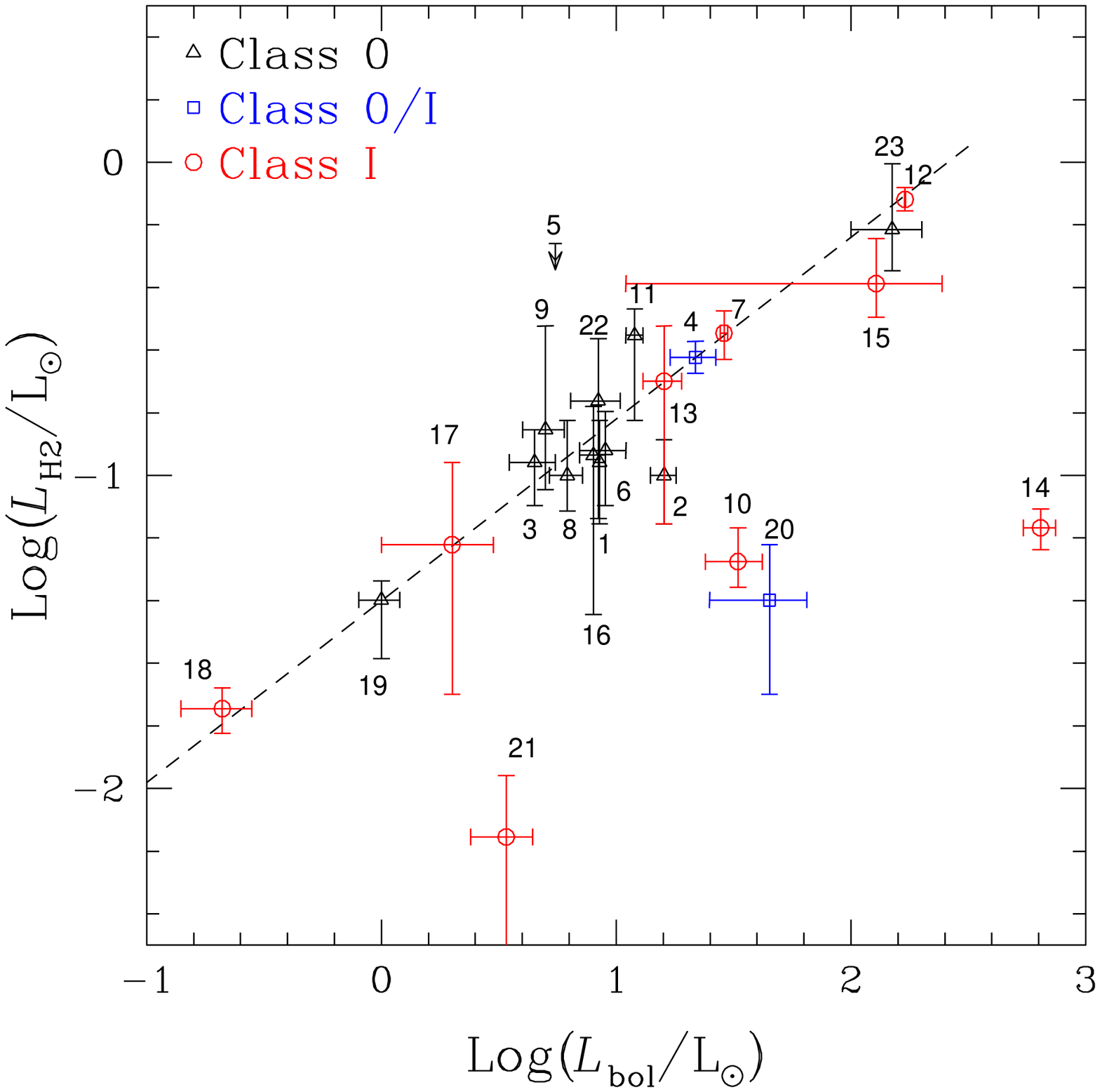}
  \caption{ Measured outflow H$_2$ luminosity (reported in Table~\ref{phys:tab}) versus the bolometric source luminosity (listed in
   Table~\ref{sample:tab}). Each object has its ID number. Outflows from Class 0 are marked with a circle, Class I with triangles and
   Class 0/I with squares. The dashed line represents the derived best-fit: 
   $log(L_{H_2}/L_{\sun}) = (0.58\pm0.06)log(L_{bol}/L_{\sun})-(1.4\pm0.04)$, computed
   without considering the four data points (ID 10, 14, 20 and 21) which fall definitively outside the relationship.
\label{lbol:fig}}
\end{figure*}

\begin{figure}
 \centering
 \includegraphics [width=9 cm]{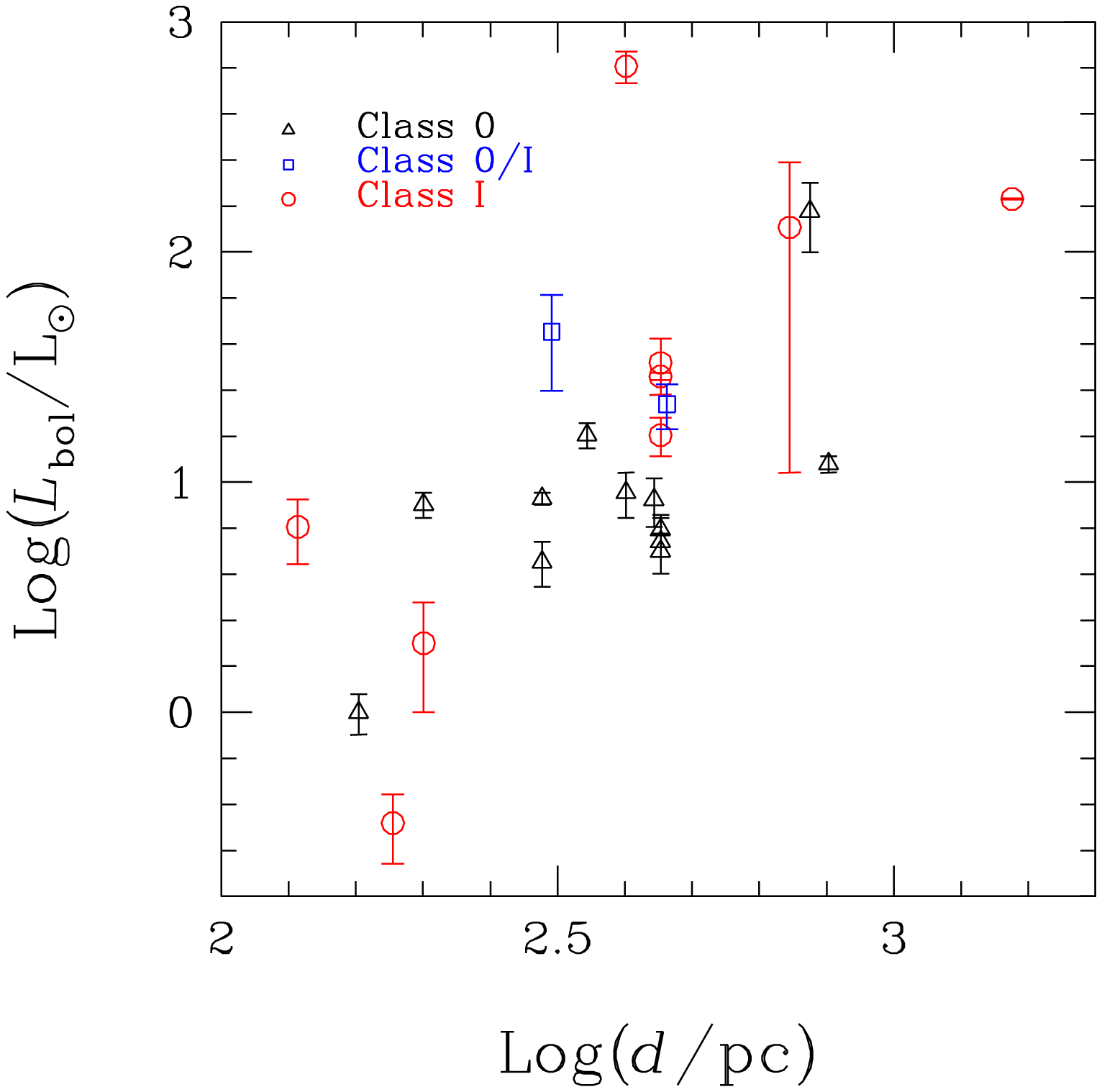}
  \caption{ Bolometric luminosity ($L_{bol}$) of the 23 sources of our sample, plotted as a function of the
            source distance from the Sun. Class 0  sources are marked with a circle, Class I with triangles and
            Class 0/I with squares. 
\label{distlbol:fig}}
\end{figure}

Since the H$_2$ emission is more easily observed from the ground than the
emission of the other major shock coolants, several works have tried
to address if a relationship between $L_{H_2}$ and $L_{\rm bol}$
also exists and depends on the protostellar evolution. Analysing a large sample
of H$_2$ jets in Orion A, \citet{stanke} found that only Class I sources 
roughly follow a correlation between $L_{H_2}$ and $L_{\rm bol}$. \citet{froebrich1}
present a similar relationship considering a sample
composed of mostly Class 0 objects. No clear correlation is found 
in this sample although a general trend is seen.
In these works the $L_{H_2}$ value is always computed from the observed
2.12\,$\mu$m flux and correction factor equal for all the flows is applied
to take into account the contribution from the other H$_2$ lines and to
correct for extinction effects. 
Here we want to address this issue again, given our more accurate determination
of $L_{H_2}$ based on the excitation conditions and $A_{\rm v}$ value for each
object. In Fig.~\ref{lbol:fig} we compare the measured outflow H$_2$
luminosity (Table~\ref{phys:tab}) versus the bolometric source luminosity 
(Table~\ref{sample:tab}), both on a logarithmic scale. Each object has
its ID number (as in Table~\ref{sample:tab}).
The greater part of the data points
lie on a straight line, showing a clear correlation
between $L_{bol}$ and $L_{H_2}$, irrispective of the Class 
of the exciting source. A few sources, i.e. three Class I and one Class 0/I,
are however displaced below this straight line, indicating that
their $L_{H_2}$/$L_{\rm bol}$ ratio is about one order of magnitude 
lower than in the other sources.
The slope follows a law of the type $L_{H_2}$ $\propto$ $L_{bol}^{0.58}$,
which closely resembles that found by \citet{shepherd} for the
mass loss rates of a sample of outflowing sources 
(i.\,e. $\dot{M_{out}} \propto L_{bol}^{0.6}$ ). However, some
caution needs to be exercised in a quantitative interpretation of
correlations involving physical quantities both dependent on
the distance. An observational bias 
could be introduced if sources at different distances are considered. 
Our source sample spans
distances ranging from 130\,pc to 1.5\,kpc and a correlation, although
with a large scatter (linear regresion gives a correlation coefficient r=0.65) is seen in a Log(D(pc)) vs Log($L_{bol}$) 
plot (see Fig.~\ref{distlbol:fig}). Given the small scatter in the correlation derived 
in Fig.~\ref{lbol:fig} (the linear regresion gives a correlation coefficient r=0.99) we can conclude that there is a
dependence of $L_{H_2}$ on $L_{bol}$, whose exact form we cannot deduce from the data.

\begin{figure}
 \centering
 \includegraphics [width=9 cm]{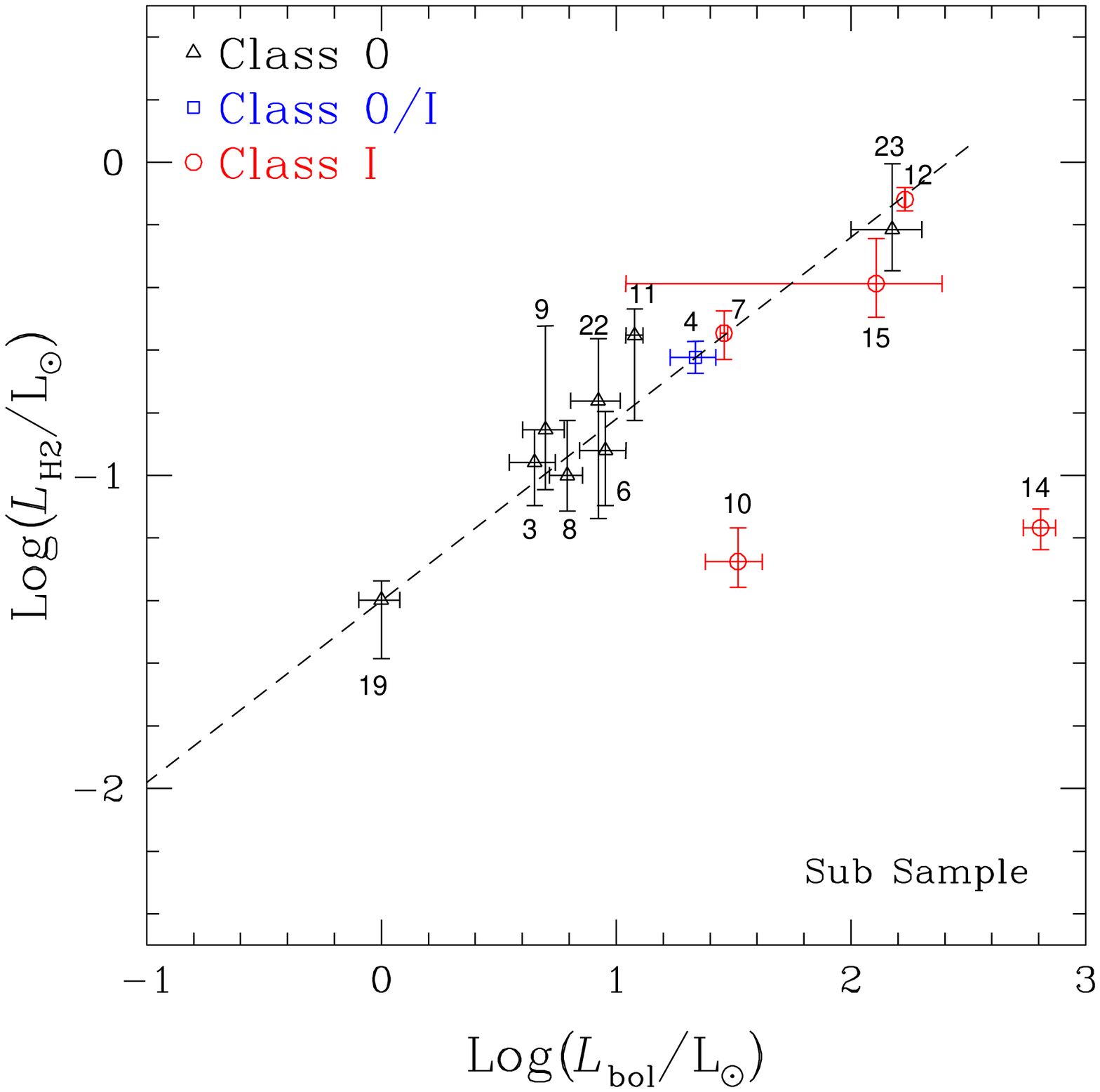}
  \caption{ Measured outflow H$_2$ luminosity (reported in Table~\ref{phys:tab}) versus the bolometric source
   luminosity (listed in Table~\ref{sample:tab}) for a sub-sample of objects, showing a clear morphology 
   (see Table~\ref{sample:tab}, col.~3). Labels are as in Fig.~\ref{lbol:fig}. The dashed line is the best-fit
   derived from the entire sample and reported in Fig~\ref{lbol:fig}.
\label{subsample:fig}}
\end{figure}

To test the accuracy of our analysis, we report in Fig.~\ref{subsample:fig} the same Log(L$_{bol}$) vs Log(L$_{H_2}$)
plot but for the sub-sample of objects showing a bipolar structure or a clear morphology
(see Sect.~\ref{sample:sec}). We find for them the same correlation obtained in the complete sample.

While Class 0 sources all show the same $L_{H_2}$/$L_{bol}$ efficiency,
the Class I sources of our sample have a less defined behaviour,
and a clear separation between the two class of sources is not
seen, at variance with the results found by \citet{paper2},
adopting the far IR line cooling as a tracer of the outflow power.

Such a different behaviour between
$L_{H_2}$ and $L_{FIR}$ could be due to the way in which the shock
luminosity is distributed among the various cooling channels,
depending on the local physical conditions. In particular, the
different critical densities of NIR H$_2$ lines ($\sim
10^{3}-10^{4}$\,cm$^{-3}$ at $T = 2000$\,K) and FIR CO and H$_2$O
lines ($10^{5}-10^{9}$\,cm$^{-3}$) may play a role. In high
density environments the H$_2$ line intensity per unit volume has already reached its
maximum LTE value and the cooling by other species (CO and H$_2$O
in particular) becomes the dominant contribution. 

In order to verify this hypothesis we have compared $L_{H_2}$ to $L_{FIR}$ 
derived in previous works \citep{paper1,paper2} for the 9 objects of our sample
where this value is available.
Although only two Class I objects are in this restricted sample, we observe a
lower ratio ($L_{H2}/L_{FIR}$) in Class 0 objects (0.9$\pm$0.7) with
respect to Class I ($\sim$13), suggesting that the density could play a 
major role in determining the cooling channels.

The plot in Fig.~\ref{lbol:fig} also shows that among our sample 
of Class I objects there is a significant number of objects ($\sim$17\%)
having an $L_{H_2}$/$L_{bol}$ ratio about an order of magnitude lower 
than the other sources. 
We have checked that for these sources
systematic errors in the derivation of $L_{H_2}$, due e.\,g. to a 
missing contribution from not considered emission knots or due to
a wrong assignment of the exciting source, are not significant.
In a few cases, it is not possible to completely exclude the occurence of 
parsec-scale outflows, which, in principle, could introduce a bias in the results.
For example, HH111 is part of a giant
outflow \citep[composed of HH113, HH311 and probably HH182][]{reipurth2,wang}
and there are no NIR data available in the literature for the remaining outflow componets.
However, even assuming that each of the other outflow components contribute as much as HH111,
the total H$_2$ luminosity would be about three times higher than the presently estimated value.
Such an effect will not appreciably change the location of this object in the Log(L$_{bol}$)
vs Log(L$_{H_2}$) plot.
It is more likely that the Class I sources with the lower $L_{H_2}$/$L_{bol}$
are more evolved objects, where $L_{bol}$ is no longer
dominated by the accretion luminosity and the mass loss 
rate is declining.
This result partially reflects the broader properties in terms of
evolution among the Class I objects, where sources with different
accretion properties are found \citep[see e.\,g.][]{nisini05}.

Most of the sources of our sample have been mapped in CO and an
estimate of the mechanical power ($L_{mech}=\frac{1}{2} \dot{M_{out}} v_{CO}^2$)
in the molecular outflow is available in the literature. 
Assuming momentum balance in the shocked region between the stellar 
jet and the ambient medium, we expect that the outflow mechanical
power should be roughly equal to the total power radiated by the
shock $L_{\rm rad}$ \citep[][]{davis}.

For those sources of our sample observed also in the far IR, we can provide
a quantitative estimate of the total shock radiated
luminosity by adding the $L_{H_2}$ and the $L_{FIR}$ due to the
other major coolant.  We find a mean ratio $L_{\rm rad}$/$L_{mech}$
around unity (1.1$\pm$0.6), which indicates that the considered
shocks may have enough power to accelerate the whole molecular ouflow.

\section{Conclusions}

We have measured the cooling and the physical properties
at NIR wavelengths of a sample of protostellar jets (23 objects),
originated by Class 0 and I low-intermediate solar mass YSOs,
presenting new spectroscopic and imaging observations of 15
objects. 
We have investigated how the derived properties are
correlated with the evolution of the jets and their
exciting sources. The main results of this work can be summarized as
follows:

\begin{itemize}
\item[-] [\ion{Fe}{ii}] emission in active H$_2$
jets has been systematically investigated to define the occurrence of
embedded ionized gas in young outflows. [\ion{Fe}{ii}] emission
spots are observed in $\sim$74\% of the investigated sample, irrespective
of the driving source class. This indicates that dissociative-shocks 
are common even in high-density embedded regions which do not show
optical HH objects.

\item[-] H$_2$ line ratios have been used to estimate the visual extinction ($A_{\rm v}$)
and average temperature of the molecular gas. When observed, the [\ion{Fe}{ii}] 
1.53/1.64\,$\mu$m ratio was used to determine the electron density ($n_{\rm e}$)
of the atomic gas component. $A_{\rm v}$ values range from $\sim$2 to $\sim$15 without
any evidence of higher extinction associated with the Class 0 flows.
The H$_2$ average gas temperatures range between $\sim$2000 and 4000\,K. In
several knots, however, a stratification of temperatures is found with
maximum values up to 5000\,K. Generally, gas components at different temperatures
are associated with the knots also showing [\ion{Fe}{ii}] emission while 
thermalized gas at a single temperature is most commonly found in knots
emitting only in molecular lines.

\item[-] We have computed the total cooling due to H$_2$ and [\ion{Fe}{ii}] 
($L_{H_2}$, $L_{FeII}$) adopting the parameters derived from the line ratio analysis
in the single knots and the H$_2$ 2.12\,$\mu$m, [\ion{Fe}{ii}] 1.64\,$\mu$m luminosities
derived from the narrow band imaging.
The determination of $L_{H_2}$ strongly depends on the local gas conditions 
and in particular on the $A_{\rm v}$ value, therefore
the often-used approximation $L_{2.12} \sim 0.1 \times L_{H_2}$ 
can be wrong by up to an order of magnitude if a proper reddening is not applyed.
\item[-] By comparing the measured outflow H$_2$ luminosity with the source 
bolometric luminosity (assumed representative of the
accretion luminosity), we find that for $\sim$83\% of the sources
there is a correlation between these two quantities, with $L_{H_2}$/$L_{bol}$$\sim$0.04. 
A small sample of four sources, however, display an
efficiency $L_{H_2}$/$L_{bol}$ lower by about an order of magnitude. 
We interpret this behaviour in terms of evolution, with the sources
which are less efficient H$_2$ emitters are more evolved than
the others. 

\item[-] We also find that there is not a clear separation in terms of 
$L_{H_2}$/$L_{bol}$ efficiency between Class 0 and Class I sources (although
the four objects with the lower $L_{H_2}$/$L_{bol}$ value are all from Class I or
intermediate Class 0/I). This partially reflects the large heterogeneity
between the evolutionary properties among Class I, which include sources
with very different accretion properties. In addition, the efficiency
of the H$_2$ cooling in a dense ambient medium, likely
characterizing the Class 0 environment, can be limited by the relatively
small critical density of H$_2$ IR lines ($\sim$10$^3$-10$^4$\,cm$^{-3}$). 
Indeed, the total cooling of flows from Class 0 sources prevalently occurs
at far IR wavelengths through CO and H$_2$O emission lines and thus can be 
underestimated considering only the H$_2$ contribution.
\item[-] On the basis of the observational experience, accumulated over
the last decade, it maybe empirically adequate to refine the concept of Class
I objects. According to the present analysis of the H$_2$ jets, we suggest to define a new Class 0.5,
which is composed (as a starting point) of objects defined as Class I YSOs \citep[according to][]{lada},
but presenting a $L_{H_2}$/$L_{bol}$ ratio similar to that of Class 0 sources (i.\,e. $\sim$0.04 or higher).
We would classify as Class 0.5
the following objects: $IRAS05173-0555$, HH26IR, $IRAS07180-2356$, $IRAS08076-3556$, \#40-3 (IRS17),
BHR71(IRS2), $IRAS12515-7641$. Obviously, this tentative classification should be confirmed with
further observations.

\end{itemize}

\begin{acknowledgements}
We are grateful to Chris J. Davis for providing H$_2$ images of VLA1623 and L1157,
to Valeri Larionov and Arkadi Arkharov for providing the HH211 spectrum. We enjoyed interesting discussions with Ren\'{e}
Liseau and Jochen Eisl\"{o}ffel.
This research has made use of NASA's Astrophysics Data System Bibliographic Services and the SIMBAD database, operated
at CDS, Strasbourg, France.
\end{acknowledgements}

\section{Appendix}

Here we give a short description of each newly analysed region and comments on the new data (NIR imaging and spectrscopy)
presented in this paper.

\subsection{L1448}

\begin{figure}
 \centering
   \includegraphics [width=6 cm] {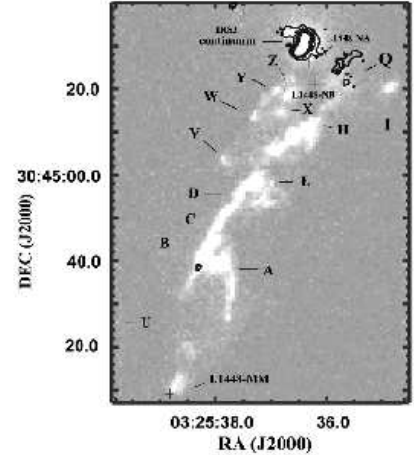}
   \caption{ L1448 outflow blue lobe: H$_2$ (2.122\,$\mu$m) image \citep[particular from][]{davismith} with the [\ion{Fe}{ii}]
   (continuum-subtracted) contours. On the IRS3 source the continuum emission from the nebula is still present as a residual.
\label{L1448:fig}}
\end{figure}

L1448 is in the Perseus dark cloud, containing several YSOs, that drive powerful outflows \citep[see e.\,g][]{davismith,eisl}.
Among them, the Class 0 L1448-MM (or C) drives an extremely high velocity outflow.
The blueshifted lobe (north-west of L1448-MM) (Fig.~\ref{L1448:fig}) is composed of (at least) three distinct curved jets
\citep{eisl}, roughly parallel, originating the millimetric source itself (the main jet) and two other embedded sources
\citep[L1448 NA, knots Z,Y,X,W,V and NB, knot Q][]{barsony}.

In this blue lobe we detected [\ion{Fe}{ii}] emission (see contours in Fig.~\ref{L1448:fig}).
Surprisingly, in the main jet knot A is not positioned at the apex of the bow-shock, but
in a region located between the west wing and knot B.

A jet-like structure, spatially coincident with knot Q, is clearly visible about 10$\arcsec$ SW of the IRS3 nebula and likely
originates L1448 NB. A second emission, a few arcsecs south of the jet, is observed.
This could be a residual of a diffuse nebulosity \citep[also observed by][]{davismith} or a real [\ion{Fe}{ii}] emission from
the main jet.

\subsection{NGC1333-I4A outflow}

NGC 1333 is a reflection nebula associated with a region of recent star formation in the Perseus molecular cloud,
with several Class 0 and I sources. Among them the IRAS 4A binary system (Class 0) drives
a well collimated bipolar outflow with several H$_2$ knots \citep{liseau,hodapp,choi}. The blue lobe (NE direction) is made
up of the H$_2$ knots ASR 57, HL 10, 11 and, possibly, HH347 \citep{bally}, and the red lobe of HL 6, 5 and 3.

Our NIR spectroscopy indicates an high visual extinction (10-15 mag) in these knots, increasing towards the source.
The high extinction explains why we do not detect some of the knots identified at mid-IR wavelengths by Spitzer 
\citep[][]{porras}.

\subsection{HH211}

HH211, discovered by \citet{mcau} in the Perseus dark cloud, is one of the smaller known outflows ($\sim$0.16\,pc).
\citet{eis2003}, with a deep H$_2$ image, observed a highly collimated jet (knot G) and counter-jet structure within an excavated
rim-brighted cavity, showing a strong continuum emission. Combined with this continuum, \citet{ocon} detected [\ion{Fe}{ii}] emission,
indicating the presence of dissociative shocks inside the outflow. In our continuum-subtracted [\ion{Fe}{ii}] image, the ionic emission on
the blue lobe (SE) clearly comes from the axis
of the bow-shock (knot I), while on the red lobe side, it is located at the end of the counter-jet (knot F) and the external bow-shocks
(knots D and C). The knot positions along with the fluxes are reported in Table~\ref{knots:tab}. As expected, the measured flux values are
lower than in \citet{ocon}, because the continuum has been removed.

\subsection{HH240/1}

The spectacular HH240/1 objects are located inside L1634 in Orion and include four pairs of symmetrical H$_2$ knots, mostly bow-shocks,
originating in the intermediate Class 0/I YSO $IRAS 05173-0555$ \citep{davis1,froeb}.
Our H$_2$ imaging and NIR spectroscopy has been presented in a previous paper \citep{paper4}.
New [\ion{Fe}{ii}] images of the outflow reveal strong compact emission from the inner bow-shocks HH240 A 
(F$_{1.64 \mu m}$ $\sim 9 \times 10^{-14} erg~s^{-1}~cm^{-2}$) and HH241 A (F$_{1.64 \mu m}$ $\sim 10^{-14} erg~s^{-1}~cm^{-2}$), 
roughly located at the apex of the bows. Unfortunately, the morphology of the iron emitting regions are poor resolved, due to
the relatively low spatial resolution of the camera (1$\arcsec$/pixel).
No other [\ion{Fe}{ii}] emission is observed along the flow.


\subsection{HH212}

Discovered by \citet{zinn}, HH212 is located in the Orion B giant
molecular cloud. Together with HH211, it represents an H$_2$ jet
prototype, visible only in the IR, with the exception of the most
distant bow-shocks, which also appears in [\ion{S}{ii}] emission
CCD images. The H$_2$ knots, centered around the source HH212--MM
or $IRAS05413-0104$, are highly symmetric and mark the periodical
outflow ejection mass. In our [\ion{Fe}{ii}] (continuum-subtracted) image, 
the knots NK1 and SK1 (nearby the source) show a strong iron component, revealed in our NIR spectra as well.
A much fainter emission is visible towards the external knots and
bow-shocks NK7, NB1 and SB1 (see Table~\ref{knots:tab}).

No other ionic component has been detected in the NIR spectra (0.95--2.5\,$\mu$m) of these knots (see Table~\ref{HH212sp1:tab}),
likely due to the high visual extinction (10--15\,mag) of the inner part of the jet.


\subsection{HH24-26 region}
\label{HH24:sec}
\begin{figure}
\centering
   \includegraphics [width=7 cm] {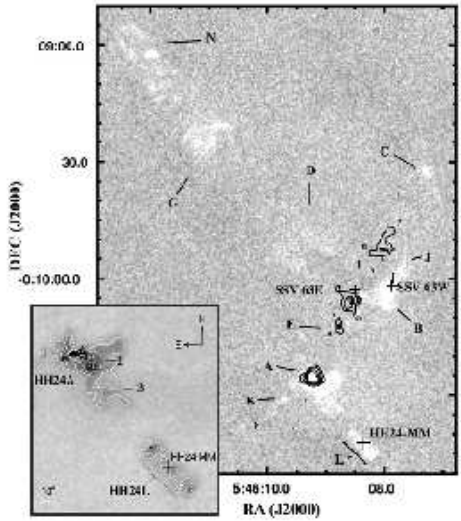}
   \caption{ HH24 region: H$_2$ (2.122\,$\mu$m) image (continuum-subtracted) with 3, 4, 5, 6$\sigma$ [\ion{Fe}{ii}]
   (continuum-subtracted) contours.
\label{HH24:fig}}
\end{figure}

This region is placed in the L1630 Orion dark cloud and exhibits a
very complex morphology due to the high star formation activity.
Three YSOs of this region are included in our sample, namely HH24--MM, HH25--MM
and HH26IRS.
HH24--MM \citep{chini1} is a Class 0 YSO, driving a highly
collimated jet \citep{bontemps} (Fig.~\ref{HH24:fig})
\citep[indicated as knot L, following][]{eis}. In the north-western
direction, roughly along the jet axis, there is the bright knot
HH24 A, but it is not clear if it is part of the same outflow or
not \citep[see e.\,g.][]{solf,eis}. The [\ion{Fe}{ii}]
(continuum-subtracted) contours in Fig.~\ref{HH24:fig} distinctly
outline a jet escaping from the YSO SSV 63E (observed through
[\ion{S}{ii}] as well) \citep[see e.\,g.][]{mundt,eis}, coincident
with knots C, E and, partially, with HH24 A. Here the brightest
iron feature matches the H$_2$ structure labelled 1 in
Fig.~\ref{HH24:fig} (lower left corner). On the contrary, knot 3
has a well defined bow-shock shape pointing backwards to HH24--MM.
Knot 2 is well aligned and could be part of the outflow too. Thus
HH24 A is probably composed of two overlaying jets, coming from
the two sources.

Knots D, G and N (see Fig.~\ref{HH24:fig}, main picture) trace another outflow, which seems to be emanated from SSV63E as well.
In this case the YSO would be a doublet, even if we do not resolve the components in our image.
Vice versa, two distinct sources compose SSV 63W, separated by about 2$\arcsec$. The northern star can be associated with
knots J and K, while the southern with knot B.

[\ion{Fe}{ii}] imaging of the HH26 region shows a strong emission ($F_{1.64 \mu m}$ $\sim 10^{-14}$\,erg\,s$^{-1}$\,cm$^{-2}$)
from knot A, whereas no iron emission is detected from HH25.

\subsection{NGC2264G}

\begin{figure*}
 \centering
   \includegraphics [width=14 cm] {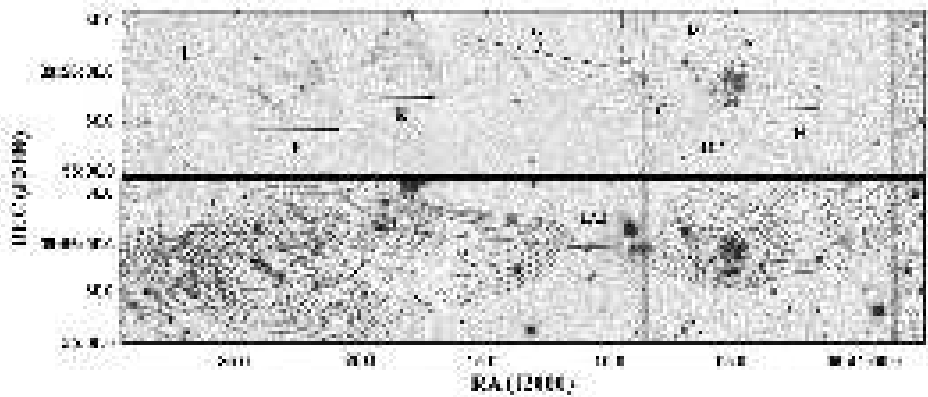}
   \caption{ NGC2264G outflow: H$_2$ (2.122\,$\mu$m) image + (continuum-subtracted). CO contours from \citet{fich}.
\label{NGC2264:fig}}
\end{figure*}

\begin{figure*}
 \centering
   \includegraphics [width=5 cm] {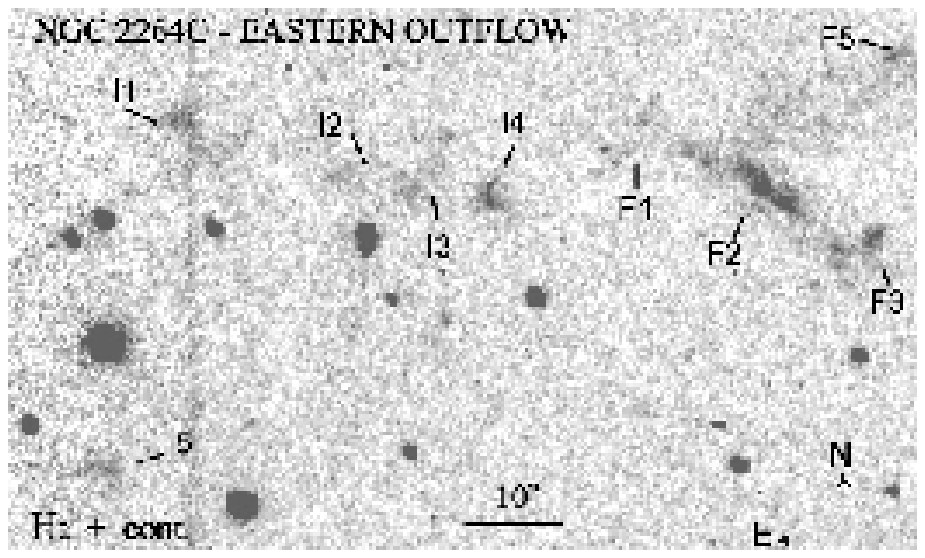}
   \includegraphics [width=5 cm] {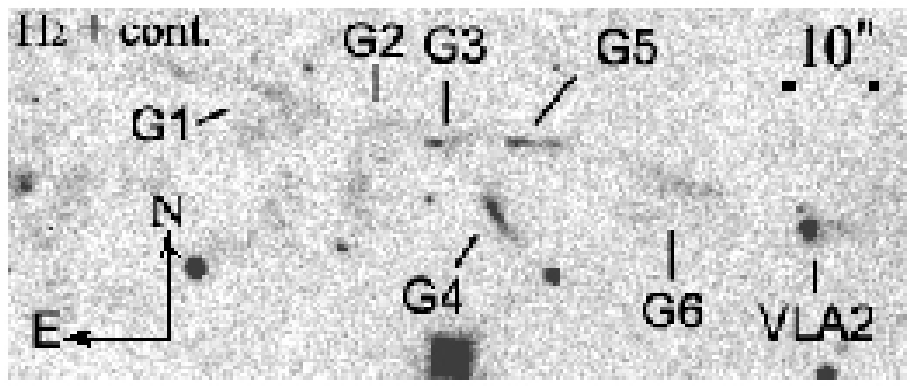}
   \includegraphics [width=5.2 cm] {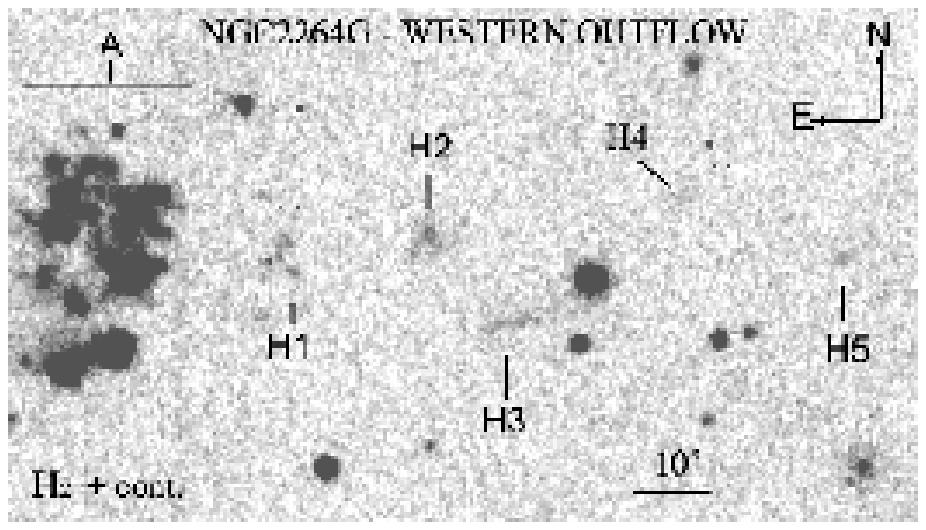} \\
   \caption{ New detected knots in the NGC2264G outflow (H$_2$ (2.122\,$\mu$m) image + continuum). {\bf Left:} Knots H in the eastern part of the outflow.
     {\bf Center:} Knots G in the central part of the outflow. {\bf Right:} Knots H in the western part of the outflow.
\label{NGC2264knot:fig}}
\end{figure*}

The high velocity outflow NGC2264G in the Mon OB1 molecular cloud (at a distance of 800\,pc) has been mapped in CO J=1-0 and 2-1 by
several authors \citep[e.\,g.][]{margulis,fich}, showing a peculiar symmetrical deflected pattern, probably due to
an abrupt change in the rotational axis.
In our image (Fig.~\ref{NGC2264:fig}), the H$_2$ emission is more extended than in previous images \citep{davis}, matching very
well the CO contours (from \citet{fich}) and appearing as a parsec scale flow (about 1.6\,pc). The H$_2$ emission in the blue lobe
(western direction) could be even wider than observed, since another CO clump has been detected just outside the image.
On the [\ion{Fe}{ii}] (continuum-subtracted) image, not reported here, we also detect a 3$\sigma$ emission (with flux
$F_{1.64 \mu m} = 1.3 \pm 0.4$\,erg\,s$^{-1}$\,cm$^{-2}$) about 5$\arcsec$ west of knot A1 and coincident with the peak of the CO outflow.
The new groups of detected knots (G, H and I), labelled following \citet{davis}, with their fluxes and positions are reported
in Table~\ref{knots:tab} and are shown in detail in Fig.~\ref{NGC2264knot:fig}.

NIR spectra of the blue lobe show only low excitation H$_2$ emission ($v \le 3$) (see Table~\ref{NGC2264Gsp:tab}).

\subsection{VLA1623}

Located in the $\rho$ Ophiuchi cloud (d$\sim$160\,pc), the VLA1623 molecular outflow \citep{andre0}
shows a very complex structure, probably because VLA1623 is a binary system \citep{looney}. H$_2$ emission was
first detected by \citet{davis} and a detailed classification can be found in \citet{gomez}.

\begin{figure*}
 \centering
   \includegraphics [width=15 cm] {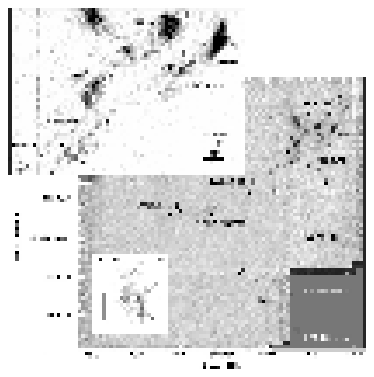}
   \caption{ VLA 1623-243 region: H$_2$ (2.122\,$\mu$m) image (continuum-subtracted). Inset (upper left corner) is an
   enlargement of the north-west part of the outflow, showing the tangential velocities and the position angles (P.A.) (with their relative
   errors) of the brightest knots. The contours are 3, 4, 5, 10 and 20$\times$ the standard
   deviation to the mean background (8$\times$10$^{-15}$\,erg\,s$^{-1}$\,cm$^{-2}$\,arcsec$^{-2}$).
   In the lower right corner, the tangential velocity and P.A. of HH313 A knots are plotted over their contours \citep{davis2}.
\label{VLA1623:fig}}
\end{figure*}

In Fig.~\ref{VLA1623:fig} we present our H$_2$ continuum-subtracted image of the region
that clearly exhibits two outflows with the typical ''S-shape" originating in the rotational axis precession of the exciting sources.
The newly detected knots (GSWC2003--17b, 14h and 14i) (see Fig.~\ref{VLA1623:fig}, upper left corner) are labelled following
the \citet{gomez} classification and their positions and fluxes are reported in Table~\ref{knots:tab}.

To better define which knots arise from the VLA1623 system, we computed the proper motions (PMs) and position
angles (P.A.) (Table~\ref{kin:tab}) for the north-western part of the outflows (see Fig.~\ref{VLA1623:fig}, upper left corner),
comparing our H$_2$ image (March 2003) with that published by \citet{davis} (April 1993) 
\citep[see][for the details on the adopted technique]{paper6}.

As suggested by \citet{eis}, we find that HH313 B is not
generated by the VLA 1623 source. Its position angle, derived from
the proper motion measurement (46$\pm$16\,km\,s$^{-1}$), is roughly 243$\degr$
(counterclockwise), hence the source could be inside the nebula
that hides the source GSS30 or it could be part of a larger flow that
includes the objects HH79 and HH711 located outside our image (M.D. Smith priv. comm.).

HH313 A is composed of two different structures (Fig.~\ref{VLA1623:fig} lower left corner).
The bow-shock \citep[knots A1, A3 and A4][]{davis2} is generated by VLA1623, showing a P.A. of 304$\degr$$\pm$8$\degr$ and
a proper motion of 60$\pm$15\,km\,s$^{-1}$, that, corrected by its inclination angle with respect to the plane of the sky 
($\sim$15$\degr$°), is in good agreement with model predictions \citep{davis2}.
The second component (knot A2) has a slower motion (46$\pm$15\,km\,s$^{-1}$) with a P.A. of 333$\degr$$\pm$24$\degr$.
From the PM analysis it is not clear if it is correlated with the bow-shock or not.

The remaining knots analysed clearly originate in the VLA
system and their slightly different P.A. follow the precession of
the two outflows.

In the large field image (Fig.~\ref{VLA1623:fig}) we also detect
two new ouflows not originating in VLA1623. The first is composed
of knots A and B, southward of HH313, symmetrically situated with
respect to ISO--OPH26, a Class II YSO \citep{bontemps2}. The
second outflow lies southward of VLA1623 and is traced by two
bow-shocks (labelled C and D), possibly generated by
$IRAS16223--2421$, located outside our image. The two bow-shocks were
also identified by \citet{khanz} (knots i and h in their field
10-01) as part of a larger outflow made up of several knots roughly
aligned with YLW 31, a Class II protostar, lying near the position
of the IRAS source.

Outflow NIR spectra (Table~\ref{HH313A:tab}-\ref{VLA1623sp3:tab}) reveal only ionic emission in HH313 A ([\ion{Fe}{ii}]
and [\ion{C}{i}]), whereas in the remaining knots only low excitation H$_2$ emission lines ($v \le 3$) are detected.

\begin{table}
\caption[]{ Tangential velocities (km\,s$^{-1}$) and position angles ($\degr$) of the studied knots in VLA1623.
    \label{kin:tab}}
\begin{center}
\begin{tabular}{ccc}
\hline \hline
knot   & $v_{tan}$ &  $P.A.$ \\
       & (km\,s$^{-1}$) & ($\degr$) \\
\hline

  HH313 A           & 60$\pm$15      & 315$\pm$12    \\
  HH313--A2        & 46$\pm$15      & 333$\pm$24     \\
  HH313 B           & 46$\pm$16      & 243$\pm$14   \\
  GSWC2003--13a    & 46$\pm$16      & 297$\pm$14   \\
  GSWC2003--13b    & 58$\pm$15      & 315$\pm$12     \\
  GSWC2003--14e    & 86$\pm$15      & 315$\pm$12     \\
  GSWC2003--17b    & 74$\pm$15      & 303$\pm$12     \\
\hline \hline
\end{tabular}
\end{center}

\end{table}

\subsection{HH54}
\begin{figure}
 \centering
   \includegraphics [width=6.5 cm] {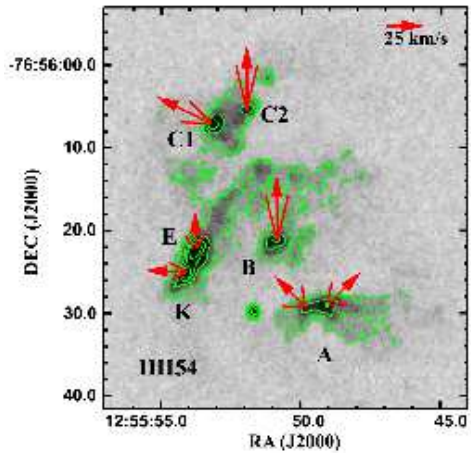}
   \caption{ HH54: H$_2$ (2.122\,$\mu$m) image (not continuum-subtracted). The arrows show the derived total velocities and the
    position angles
    (P.A.) (with their relative errors) of the brightest knots. The contours are 3, 5, 10, 15 and 20$\times$ the standard
   deviation to the mean background (5$\times$10$^{-15}$\,erg\,s$^{-1}$\,cm$^{-2}$\,arcsec$^{-2}$).
\label{hh54:fig}}
\end{figure}

HH54 is situated in a low mass star-forming region of the Chamaleon dark clouds at a distance of about 200\,pc \citep{hug}.
There are at last two driving sources candidates for the outflow, IRAS 12522-7640 and IRAS 12515-7641 \citep{knee}. The first
lies inside the Herbig Haro object, roughly near knot K position and the second lies westward.
Our study of the proper motion does not clarify this uncertainty (Fig.~\ref{hh54:fig}, see values in Table~\ref{kinhh54:tab}). 
THe HH54 knots move roughly in the NNE direction, different from the IRAS 12515-7641 position. Moreover, the inferred outflow
inclination angle with respect to the sky
($\sim 15\degr$), obtained combining both tangential and radial velocities \citep[from][]{paper8}, indicates that the source should be
located firther south than IRAS 12522-7640, where HH54 X and Y are observed, but in this region no other YSOs are detected.
Although P.M. analysis does not reveal the driving source, following \citet{knee}, we assume that HH54 is the red 
lobe of the outflow containing HH52 and HH53 and that IRAS 12515-7641 is the exciting source.

\begin{table}
\caption[]{ Tangential and total velocities (km\,s$^{-1}$), position angles ($\degr$) of the studied knots in HH54.
    \label{kinhh54:tab}}
\begin{center}
\begin{tabular}{cccc}
\hline \hline
knot   & $v_{tan}$ & $v_{tot}$ & $P.A.$ \\
       & (km\,s$^{-1}$) & km\,s$^{-1}$) &($\degr$) \\
\hline

  HH54 A    & 36$\pm$18  & 37$\pm$18    & 315$\pm$31 \& 45$\pm$31$^*$    \\
  HH54 B    & 50$\pm$18  & 52$\pm$18    & 0$\pm$16     \\
  HH54 C1   & 56$\pm$18  & 58$\pm$18    & 63$\pm$15   \\
  HH54 C2   & 50$\pm$18  & 52$\pm$18    & 0$\pm$16   \\
  HH54 E    & 25$\pm$18  & 28$\pm$18    & 0$\pm$31   \\
  HH54 K    & 25$\pm$18  & 28$\pm$18    & 90$\pm$20   \\
\hline \hline
\end{tabular}
\end{center}
Notes: $^*$ HH54 A has two peaks (estward and westward, see Fig.~\ref{hh54:fig}), showing same P.M.s but different P.A.s.\\
\end{table}
\begin{figure*}
 \centering
   \includegraphics [width=15 cm] {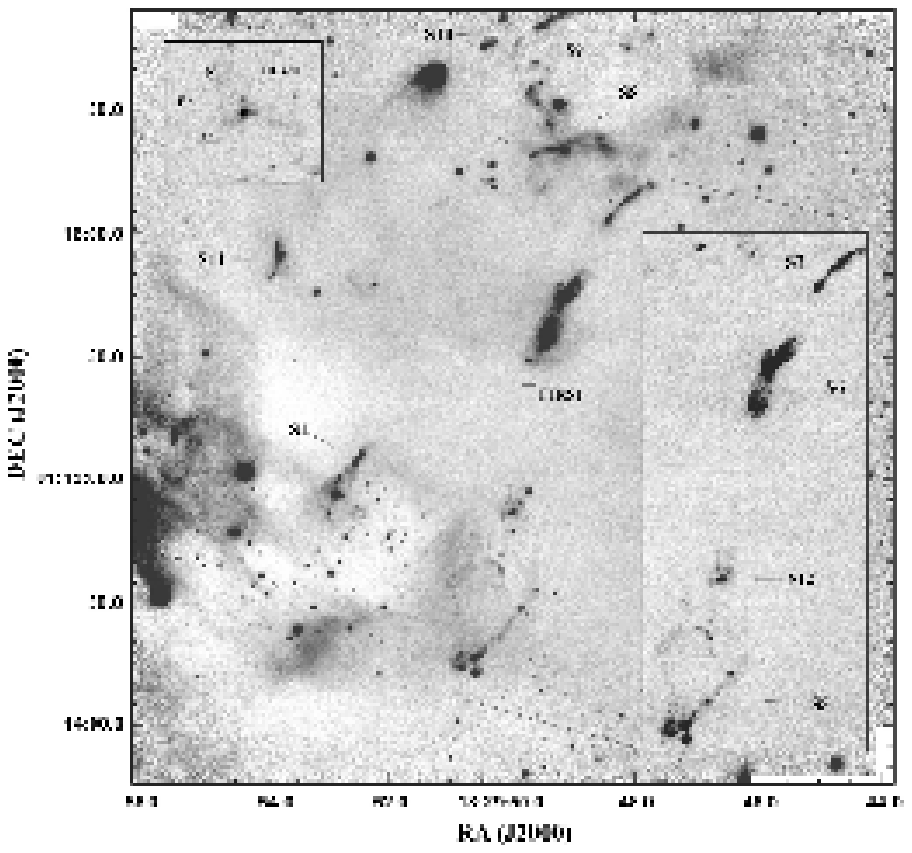}
   \caption{ $IRAS18273+0113$ region: H$_2$ (2.122\,$\mu$m) image (not continuum-subtracted) (main picture).
     On the lower right panel, we show the H$_2$-K image of the FIRS1 main body outflow, on the upper left
     the H$_2$-K image of HH460, located NW outside the main picture and possibly part of the outflow.
\label{serp:fig}}
\end{figure*}

\subsection{IRAS18273+0113}

$IRAS18273+0113$ \citep[FIRS1 or SMM1,][]{harvey}, in the crowded Serpens star-forming region, powers
a bipolar outflow, well traced at NIR wavelengths \citep[e.\,g.][]{herbst,eiroa}.
In Fig.~\ref{serp:fig}, the H$_2$ image (not continuum-subtracted) shows manifold jets, originated by the
several millimetric sources of the region. Following the \citet{herbst} notation, the outflow driven by FIRS1 coincides with knots S5
in the red lobe and with S6 and S7 in the blue lobe.
In the H$_2$ continuum-subtracted image (Fig.~\ref{serp:fig} lower right panel), S5 exhibits a clear bow-shock structure and is followed by
another shocked feature labelled S12.

The sub-mm CO studies \citep{davis3} shows that the outflow is
much more extended and that knots HH460 (located NW of the source,
see Fig.~\ref{serp:fig} upper left panel) and, possibly, HH459
\citep[][]{ziener,davis3}
could be part of it. Probably a deep H$_2$ image of the SE and NW
parts of the flow could reveal other molecular emissions.

Conversely, a deep [\ion{Fe}{ii}] narrow band image does not manifest any iron emission, within an upper limit flux of
$F_{1.64 \mu m}$=5.3$\times$10$^{-16}$\,erg\,s$^{-1}$\,cm$^{-2}$\,arcsecs$^{-2}$.

\subsection{R CrA--HH99}

\begin{figure*}
 \centering
   \includegraphics [width=14 cm] {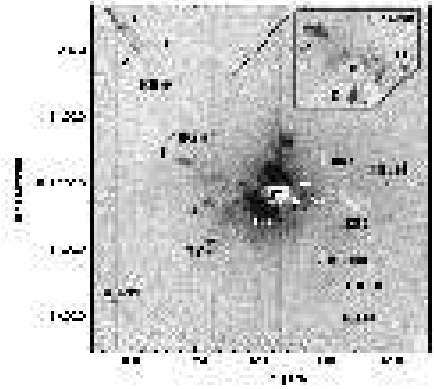}
   \caption{H$_2$ (2.122\,$\mu$m) (continuum-subtracted) image of the RCrA region. The arrow shows the tangential velocity and
    the position angle (P.A.) (with its relative error) of HH99. 
    The newly detected knots (upper right corner) are labelled following \citet{davis3}.
\label{rcra:fig}}
\end{figure*}

R CrA dark cloud is one of the star forming regions closest to the Sun and
harbours several YSOs, outflows and HH objects, such as HH99.
The exciting source of HH99 is still uncertain. \citet{hart} tentatively indicated
HH100IR as the driving source of an outflow including HH100 (blue lobe)
and HH99 (red lobe). With NIR imaging, \citet{wilking} as well
as \citet{davis3} suggest IRS 9 or the source R CrA itself.
From [\ion{S}{ii}] large field imaging, \citet{wang1}
indicate IRS 6 as a possible candidate, but the protostar appears
too evoluted to have given birth to such a jet \citep{nisini05}. 
Our proper motion analysis of HH99 (see Fig.~\ref{rcra:fig}), obtained from NTT and VLT H$_2$ images ($\Delta$t$\sim$6~yr),
indicates a P.A. of 39$\degr$$\pm$6$\degr$ and tangential velocity of 118$\pm$14\,km\,s$^{-1}$ (at a distance
of 130\,pc), excluding both IRS 6 and 9 as exciting sources. The Class I IRS 7 (here assumed as the driving source) 
or even a more embedded object of this system \citep[see e.\,g.][]{nutter,hamaguchi} are the best candidates.

\subsection{L1157}
\begin{figure}
 \centering
   \includegraphics [width=6 cm] {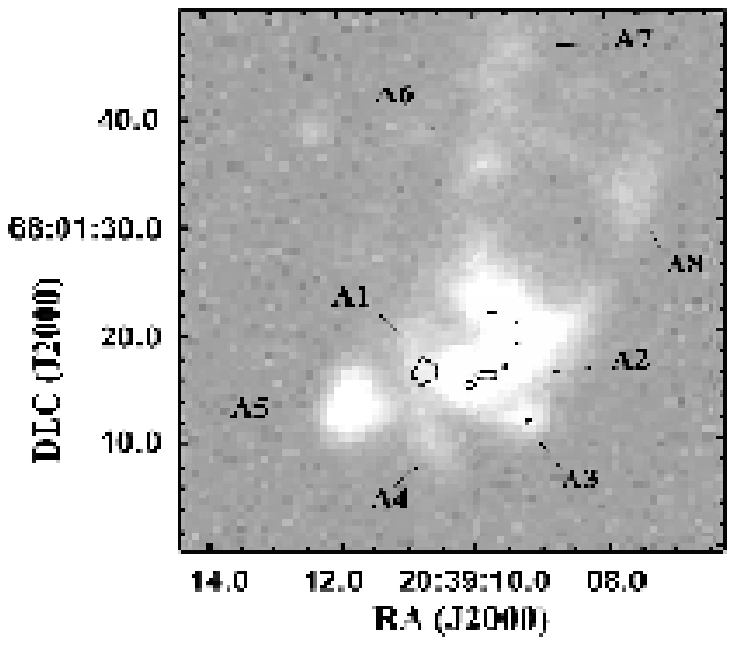}
   \caption{ L1157 knot A: H$_2$ (2.122\,$\mu$m) image \citep[from][]{davis} with  a 3$\sigma$ [\ion{Fe}{ii}] (continuum-subtracted)
   contour.
\label{L1157:fig}}
\end{figure}

In our [\ion{Fe}{ii}] image we detect emission only in the blue
lobe outflow. In Fig.~\ref{L1157:fig} \citep[from][]{davis}, we
show the bright knot A, located in the blue lobe of the outflow, 
with the iron line (continuum-subtracted)
matching the substructures labelled as A1 and 2 in \citet{davis},
lying on the jet axis with a total measured flux of
$F_{1.64 \mu m}$=1.1$\pm$0.3 $\times$10$^{-14}$\,erg\,s$^{-1}$\,cm$^{-2}$.

\subsection{IC1396N}

$IRAS21391+5802$ is located inside the bright rimmed globule IC 1369--N in the Cep OB2
association, a very young and active intermediate mass star-forming region. The region around the IRAS source
harbours three YSOs (namely BIMA 1, 2 and 3) \citep{codella,beltran} and the first two generate an outflow.

\begin{figure*}
 \centering
   \includegraphics [width=15 cm] {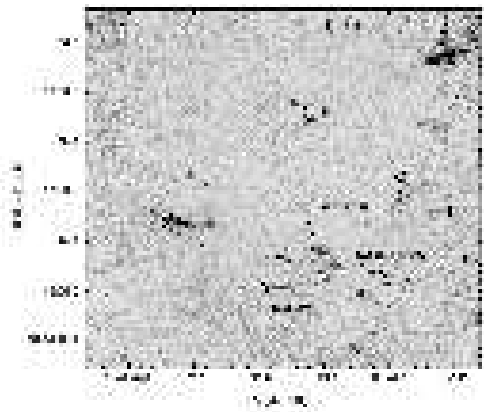}
   \caption{ IRAS21391+5802: H$_2$ (2.122\,$\mu$m) image (continuum-subtracted) with superimposed 3 and 4 $\sigma$ [\ion{Fe}{ii}]
   contours (white on the H$_2$ emission and black on the background). We label four newly detected knots 
   (N, O, P and Q, see Table~\ref{knots:tab}) in the image, following the nomenclature of \citet{nisini1} and \citet{reip}.
\label{ic1396:fig}}
\end{figure*}

Fig.~\ref{ic1396:fig} (H$_2$ continuum-subtracted image) shows the
complex morphology of this region, where at least four outflows
can be detected \citep[see][for details]{nisini1,reip}. 
BIMA 2, the most massive object associated with the IRAS
source, gives birth to a spectacular outflow, roughly oriented
east--west \citep{codella,beltran}. Knots B, A ,M and HH593 are
spatially coincident with the outflow's red lobe, but their
location would trace the outflow cavity rather than
the jet axis itself. Even if there was a better alignment between
the knots and the source, the hypothesis that BIMA 3 has generated
knots A and B \citep{reip} seems less likely, because
no molecular outflow from the object has been detected in centimetric and millimetric
observations \citep{codella,beltran}. The blue lobe
shows less H$_2$ emission (knots K and Q), possibly because the
local extinction is higher.

A strong continuum emission is present in both [\ion{Fe}{ii}] and H$_2$ images, probably due to
the illumination of the nebula by the O6 star HD206267.
The 3$\sigma$ contours in Fig.~\ref{ic1396:fig} (superimposed on the 2.12\,$\mu$m line) show that the [\ion{Fe}{ii}]
emission (continuum-subtracted) mainly originates from knots A and HH593 and near the IRAS source, tracing the higher excited regions.
A further [\ion{Fe}{ii}] detection is observed towards HH777 IRS YSO.

\begin{table*}
\caption[]{ H$_2$ (2.122$\mu$m) and [\ion{Fe}{ii}] (1.644$\mu$m)
 fluxes and positions of the new knots detected toward the investigated outflows.
    \label{knots:tab}}
\begin{center}
\begin{tabular}{cccccccccc}
\hline \hline
OUTFLOW & KNOT--ID &$F$(2.12$\mu$m) $\pm$ $\Delta$$F$& $F$(1.64$\mu$m) $\pm$ $\Delta$$F$ &\multicolumn{3}{c}{$\alpha$(2000.0)}& \multicolumn{3}{c}{$\delta$(2000.0)}\\

  &   & ($10^{-15}$erg\,s$^{-1}$\,cm$^{-2}$)  & ($10^{-15}$erg\,s$^{-1}$\,cm$^{-2}$) & ($^{h}$ & $^{m}$ & $^{s}$) & ($\degr$ & $\arcmin$ & $\arcsec$) \\
\hline \hline
L1448   &      A-FeII   & $\cdots$      &  3.7$\pm$1.2      & 03&25&38.3 &30&44&34.2 \\
L1448   &      Q--jet   & $\cdots$      &  26$\pm$7         & 03&25&35.7 &30&45&19.9 \\
HH211   &     C       & $\cdots$      &  3$\pm$1        & 03&43&53.4 &32&01&14.2 \\
HH211   &     D       & $\cdots$      &  12$\pm$2       & 03&43&53.9 &32&01&04.5 \\
HH211   &     F       & $\cdots$      &  4.7$\pm$0.7        & 03&43&55.0 &32&01&02.7 \\
HH211   &     I       & $\cdots$      &  11$\pm$2       & 03&43&59.4 &32&00&35.0 \\
HH212   &     NK1       & $\cdots$      &  6.3$\pm$1.0      & 05&43&51.5 &-01&02&48.6 \\
HH212   &     SK1       & $\cdots$      &  5.7$\pm$1.0      & 05&43&51.3 &-01&02&58.7 \\
HH212   &     NB1       & $\cdots$      &  2.8$\pm$1.0      & 05&43&52.3 &-01&02&16.2 \\
HH212   &     SB1       & $\cdots$      &   $<$ 1                 & 05&43&50.3 &-01&03&27.3 \\
HH24    &     A       &  $\cdots$      &  20$\pm$2      & 05&46&09.2 &-00&10&25.1 \\
HH24    &     C       &  $\cdots$      &  27$\pm$3      & 05&46&00.0 &-00&09&51.9 \\
HH24    &     E       &  $\cdots$      &  8$\pm$2       & 05&46&08.6 &-00&10&05.7 \\
HH24    &     D       &  17$\pm$2    &  $\cdots$            & 05&46&09.3 &-00&09&49.5 \\
HH24    &     G       &  44$\pm$2    &  $\cdots$        & 05&46&11.0 &-00&09&22.0 \\
HH24    &     N       &  38$\pm$2    &  $\cdots$        & 05&46&12.0 &-00&09&02.0 \\
NGC2264G &  G1       & 2.2$\pm$0.4      &   $\cdots$     & 06&41&15.3 &09&56&14.9 \\
NGC2264G &  G2       & 2.3$\pm$0.5      &   $\cdots$    & 06&41&14.4 &09&56&10.9 \\
NGC2264G &  G3       & 2.4$\pm$0.3      &   $\cdots$         & 06&41&14.0 &09&56&09.2 \\
NGC2264G &  G4       & 5.5$\pm$0.3      &   $\cdots$     & 06&41&13.5 &09&56&00.9 \\
NGC2264G &  G5       & 4.1$\pm$0.3       &   $\cdots$    & 06&41&13.3 &09&56&09.0 \\
NGC2264G &  G6       & 2.8$\pm$0.4      &   $\cdots$         & 06&41&12.0 &09&56&04.2 \\
NGC2264G &  H1       & 4.4$\pm$0.4      &   $\cdots$     & 06&41&03.8 &09&55&53.6 \\
NGC2264G &  H2       & 3.5$\pm$0.5       &   $\cdots$    & 06&41&02.8 &09&55&56.1 \\
NGC2264G &  H3       & 3$\pm$1      &   $\cdots$         & 06&41&02.2 &09&55&46.8 \\
NGC2264G &  H4       & 1.5$\pm$0.5      &   $\cdots$        & 06&41&01.1 &09&55&59.2\\
NGC2264G &  H5       & 1.5$\pm$0.5      &   $\cdots$        & 06&41&00.0 &09&55&53.5\\
NGC2264G &  I1       & 6.0$\pm$0.5      &   $\cdots$        & 06&41&28.1 &09&55&53.5 \\
NGC2264G &  I2       & 1.8$\pm$0.3   &   $\cdots$        & 06&41&27.0 &09&55&47.7 \\
NGC2264G &  I3       & 3.5$\pm$0.4      &   $\cdots$        & 06&41&26.5 &09&55&46.4 \\
NGC2264G &  I4       & 5.8$\pm$0.3      &   $\cdots$        & 06&41&26.0 &09&55&45.7 \\
NGC2264G &  I5       & 2.0$\pm$0.5      &   $\cdots$        & 06&41&28.6 &09&55&17.1 \\
VLA1623 & GSWC2003--14h & 17$\pm$4      &  $\cdots$         &16&26&20.6&-24&23&15.6 \\
VLA1623 & GSWC2003--14i & 4$\pm$1       &  $\cdots$         &16&26&19.8&-24&23&19.1 \\
VLA1623 & GSWC2003--17b & 37$\pm$8      &  $\cdots$         &16&26&24.0&-24&24&09.7 \\
OPH--ISO &     A        & 1.6$\pm$0.5   &  $\cdots$         &16&26&20.0&-24&24&03.2 \\
OPH--ISO &     B        & 2.0$\pm$0.5   &  $\cdots$         &16&26&18.0&-24&24&27.2 \\
IRAS16233&     C        & 17$\pm$4      &  $\cdots$         &16&26&25.2&-24&27&34.1 \\
IRAS16233&     D        & 36$\pm$8      &  $\cdots$         &16&26&28.4&-24&27&04.3 \\
IRAS18273    &     S12     & 5.4$\pm$1.0   &  $\cdots$         &18&29&50.2&01&14&51.6 \\
HH593    &      & $\cdots$      & 13$\pm$4          &21&40&44.9&+58&16&09.5 \\
IC1396N  &     A        & $\cdots$      & 21$\pm$5          &21&40&44.9&+58&16&20.4 \\
IC1396N  &     N        & 6.2$\pm$0.8      & $\cdots$      &21&40&47.2&+58&16&01.7 \\
IC1396N  &     0        & 4.3$\pm$0.9      & $\cdots$      &21&40&53.5&+58&17&01.6 \\
IC1396N  &     P        & 19.8$\pm$1.4      & $\cdots$      &21&40&33.7&+58&17&55.3 \\
IC1396N  &     Q        & 10.1$\pm$1.5      & $\cdots$      &21&40&41.7&+58&15&54.2 \\
IC1396N  &     R        & 14$\pm$2      & $\cdots$          &21&40&40.2&+58&15&10 \\
\hline \hline
\end{tabular}
\end{center}
\end{table*}

\begin{table*}
\caption[]{Physical parameters of the newly observed knots derived through spectroscopy.
    \label{phys_knots:tab}}
\begin{tabular}{ccccc}
\hline\\[-5pt]
knot  (object)   & Area (10$^{-10}$~sr)$^{*}$ &   $T(K)$  & $A_{\rm v}$ (mag) & $n_{\rm e}$ (cm$^{-3}$)  \\
\hline\\[-5pt]
knot 5 (NGC1333I-4A) & 3.14  &  2150$\pm$150     &  8$\pm$3    &  $\cdots$ \\
ASR57 (NGC1333I-4A) &  1.86  &  2650$\pm$150     &  12$\pm$3     &  $\cdots$ \\
i+j (HH211)        &  7.3     & 2650$\pm$150     &  8$\pm$3     &  $\cdots$ \\
NB1 (HH212)        &  1.4     & 2650$\pm$150$^b$    &  4$\pm$1  &  $\cdots$ \\
SB3 (HH212)        &   2      & 2650$\pm$150$^b$    &  4$\pm$1  &  $\cdots$ \\
NK1 (HH212)        &   1.36   & 2800$\pm$150$^b$ &  11$\pm$2$^a$  &  $1.6 \times 10^4$ \\
SK1 (HH212)        &  1.39    & 2950$\pm$180$^b$ &  12$\pm$2$^a$  &  $10^4$-$10^5$ \\
NK2 (HH212)        &  0.62    & 2260$\pm$160     &  12$\pm$3       &  $\cdots$ \\
NK4 (HH212)        &  0.8     & 2300$\pm$100     &  12$\pm$3  &  $\cdots$ \\
NK7 (HH212)        &  0.92    & 2300$\pm$100     &  12$\pm$3  &  $\cdots$ \\
NB2 (HH212)        &  1.4     & 2600$\pm$200       & 8$\pm$3 &   $\cdots$\\
SK2 (HH212)        &   0.6    & 2100$\pm$200       & 12$\pm$3   &   $\cdots$\\
SK4 (HH212)        &   0.47     & 1650$\pm$650     & 12$\pm$3   &   $\cdots$\\
SK5 (HH212)        &   0.43     & $\cdots$         & 12$\pm$3   &   $\cdots$\\
SB2 (HH212)        &   1.4      & 2400$\pm$200     &  8$\pm$3  &   $\cdots$\\
A2 (NGC2264G)      &   0.68     & 2500$\pm$300         & 5$\pm$3  &   $\cdots$\\
A3 (NGC2264G)      &   0.88     & 2350$\pm$250         &  5$\pm$3  &   $\cdots$\\
A7 (NGC2264G)      &   1.08     & 2450$\pm$350         & 5$\pm$3 &   $\cdots$\\
C1 (NGC2264G)      &   0.75     & 2500$\pm$300         & 5$\pm$3  &   $\cdots$\\
C2 (NGC2264G)      &   0.60     & 2500$\pm$300         & 5$\pm$3  &   $\cdots$\\
HH313 A        &   2.59     & 3130$\pm$130$^b$  & 8$\pm$2 12$\pm$2$^a$ & 1-6 $\times 10^4$ \\
HH313 B            &   3.01     & 2760$\pm$150$^b$  & 8$\pm$3     &   $\cdots$  \\
GSWC2003-13a       &   2.19     & 2500$\pm$140     & 8$\pm$4   &       $\cdots$         \\
GSWC2003-13b       &   1.01      & 2200$\pm$130     & 8$\pm$4   &     $\cdots$           \\
GSWC2003-14a       &   1.5       & 2200$\pm$130     & 8$\pm$4   &     $\cdots$           \\
GSWC2003-14c+14d   &   4.02      & 2200$\pm$130     & 8$\pm$4   &     $\cdots$           \\
GSWC2003-14e       &   1.01      & 2800$\pm$140     & 8$\pm$4   &     $\cdots$           \\
GSWC2003-14f       &   1.28      & 2370$\pm$250     & 8$\pm$2   &     $\cdots$           \\
GSWC2003-14g       &   1.70      & 2800$\pm$240     & 8$\pm$2   &     $\cdots$           \\
GSWC2003-14h       &   3.31      & 3000$\pm$400     & 8$\pm$2   &     $\cdots$           \\
GSWC2003-18        &   1.90      & 2500$\pm$300     & 12$\pm$3   &     $\cdots$           \\
GSWC2003-20b        &  1.04      & 2250$\pm$300     & 12$\pm$3   &     $\cdots$           \\
GSWC2003-20c        &  1.38       & 2250$\pm$300     & 12$\pm$3   &     $\cdots$           \\
GSWC2003-20d        &  2          & 2250$\pm$300     & 12$\pm$3   &     $\cdots$           \\
GSWC2003-20e        &  0.82       & 2200$\pm$200     & 12$\pm$3   &     $\cdots$           \\
GSWC2003-20f        &  1.64       & 2250$\pm$150     & 12$\pm$3   &     $\cdots$           \\
S6 (serpens)        &  2.18       & 2200$\pm$200     & 8$\pm$3      &     $\cdots$   \\
S7 (serpens)        &  1.71       & 2600$\pm$300     & 8$\pm$3     &     $\cdots$   \\
A1 (L1157)         &   1.78      & 3050$\pm$120$^b$ &   $<$ 2      &     $\cdots$   \\
A2 (L1157)         &  2.01        & 3070$\pm$140$^b$ &  $<$ 2      &     $\cdots$   \\
A4 (L1157)         &  0.76        & 2090$\pm$120 &   2$\pm$1  &     $\cdots$   \\
A8 (L1157)         &  0.89        & 2650$\pm$300 &   2$\pm$1  &     $\cdots$   \\
C1 (L1157)         &  1.56        & 2870$\pm$130$^b$ &  $<$ 2      &     $\cdots$   \\
C2 (L1157)         &  2.20        & 2600$\pm$130$^b$ &  $<$ 2      &     $\cdots$   \\
C4 (L1157)         &  1.22        & 2460$\pm$100$^b$ &   2$\pm$1  &     $\cdots$   \\
D3 (L1157)         &  3.64        & 2500$\pm$300 &   2$\pm$1  &     $\cdots$   \\
A (IC1396N)         & 2.88        & 2740$\pm$140 &   10$\pm$5  &     $\cdots$   \\
B (IC1396N)         & 1.02        & 2400$\pm$140 &   10$\pm$5  &     $\cdots$   \\
\hline\\[-5pt]
\\
\end{tabular}\\
Notes: 
$^{*}$Area (length $\times$ slit aperture) through which the fluxes in Tables  9-21 have been computed.\\
Temperature and extinction are derived from H$_2$ lines. We report a label ($^a$) when $A_{\rm v}$ has 
been obtained from [\ion{Fe}{ii}]. \\
$^b$ Knots that show a stratification in temperature. An average T is reported.\\
\end{table*}


\begin{table*}
\caption[]{Observed lines in NGC1333I-4A outflow knots 5 and ASR57
    \label{NGC1333sp:tab}}
\begin{tabular}{cccc}
\hline\\[-5pt]
Term&  $\lambda$($\mu$m) & \multicolumn{2}{c}{$F\pm\Delta~F$(10$^{-15}$erg\,cm$^{-2}$\,s$^{-1}$)}\\
\hline\\[-5pt]
H$_2$ Lines          &          &   knot5           &    ASR57             \\
\hline\\[-5pt]
~1--0 S(9)           & 1.688    &  $\cdots$      & 5.0$\pm$0.9          \\
~1--0 S(8)           & 1.715    &  $\cdots$      & 3.7$\pm$1.0       \\
~1--0 S(7)           & 1.748    & 7.5$\pm$0.7    & 24.1$\pm$0.9          \\
~1--0 S(6)           & 1.788    & 6.8$\pm$1.4    & 19.5$\pm$1.0          \\
~1--0 S(3)           & 1.958    &  52$\pm$5      & 105$\pm$3             \\
~2--1 S(4)           & 2.004    &  $\cdots$      & 3.8$\pm$1.0           \\
~1--0 S(2)           & 2.034    & 15.0$\pm$1.2   & 50.0$\pm$0.8          \\
~3--2 S(5)           & 2.066    &  $\cdots$      & 2.0$\pm$0.6           \\
~2--1 S(3)           & 2.073        & 4.1$\pm$1.3    & 17.8$\pm$0.6          \\
~1--0 S(1)           & 2.122    & 40.6$\pm$0.6   & 130.0$\pm$0.6         \\
~2--1 S(2)           & 2.154    &  $\cdots$      & 6.8$\pm$0.5           \\
~3--2 S(3)           & 2.201    &  $\cdots$      & 5.6$\pm$0.6            \\
~1--0 S(0)           & 2.223    &  9.6$\pm$0.9   & 33.5$\pm$0.8          \\
~2--1 S(1)           & 2.248    &  6.6$\pm$0.5   & 18.4$\pm$0.7          \\
~2--1 S(0)           & 2.355    &  $\cdots$      & 6.0$\pm$1.5           \\
~3--2 S(1)           & 2.386    &  $\cdots$      & 5.0$\pm$1.6          \\
~1--0 Q(1)           & 2.407    &   43$\pm$5     &  93$\pm$5        \\
~1--0 Q(2)           & 2.413    &   20$\pm$5     &  47$\pm$5             \\
~1--0 Q(3)           & 2.424    &   45$\pm$5     &  84$\pm$5             \\
~1--0 Q(4)           & 2.437    &   $\cdots$     &  64$\pm$5           \\
~1--0 Q(5)           & 2.455    &   $\cdots$     &  49$\pm$5           \\

\hline\\[-5pt]
\\
\end{tabular}\\
Notes: $^{a}$S/N between 2 and 3.\\
\end{table*}


\begin{table*}
\caption[]{ Observed lines in HH211 outflow: knots i and j
    \label{HH211sp:tab}}
\begin{tabular}{ccc}
\hline\\[-5pt]
Term&  $\lambda$($\mu$m) & $F\pm\Delta~F$(10$^{-14}$erg\,cm$^{-2}$\,s$^{-1}$)\\
\hline\\[-5pt]
H$_2$ Lines          &           &    I + J                 \\
\hline\\[-5pt]
~1--0 S(9)           & 1.688     & 1.6$\pm$0.3          \\
~1--0 S(8)           & 1.715     & 1.2$\pm$0.3               \\
~1--0 S(7)           & 1.748     & 7.2$\pm$0.3           \\
~1--0 S(6)           & 1.788     & 4.6$\pm$0.3           \\
~1--0 S(3)           & 1.958     & 51$\pm$2              \\
~2--1 S(4)           & 2.004     &  $\cdots$             \\
~1--0 S(2)           & 2.034     & 13.5$\pm$0.3          \\
~3--2 S(5)           & 2.066     & 0.6$\pm$0.3$^a$           \\
~2--1 S(3)           & 2.073     & 3.4$\pm$0.3       \\
~1--0 S(1)           & 2.122     & 35.2$\pm$0.3          \\
~2--1 S(2)           & 2.154     & 2.0$\pm$0.3           \\
~3--2 S(3)           & 2.201     & 1.3$\pm$0.3        \\
~1--0 S(0)           & 2.223     & 7.5$\pm$0.3           \\
~2--1 S(1)           & 2.248     & 3.6$\pm$0.3           \\
\hline\\[-5pt]
[{\ion{Fe}{ii}}] lines           &                       &          \\
\hline\\[-5pt]
~[{\ion{Fe}{ii}}]\,$a^4\!D_{7/2}-a^4\!F_{9/2}$  & 1.644  &1.1$\pm$0.3    \\
\hline\\
Notes: $^{a}$S/N between 2 and 3.\\
\end{tabular}\\
\end{table*}


\begin{table*}
\caption[]{ Observed lines in HH212 outflow
    \label{HH212sp1:tab}}
\begin{tabular}{cccc}
\hline\\[-5pt]
Term&  $\lambda$($\mu$m) & \multicolumn{2}{c}{$F\pm\Delta~F$(10$^{-15}$erg\,cm$^{-2}$\,s$^{-1}$)}\\
\hline\\[-5pt]
H$_2$ Lines          &          &   HH212 NB1   &  HH212 SB3   \\
\hline\\[-5pt]
~2--0 S(7)           & 1.064    &  2.2$\pm$0.6  &  3$\pm$1  \\
~2--0 S(5)               & 1.085    &  2.2$\pm$0.5  &  2.7$\pm$0.6  \\
~2--0 S(3)           & 1.117    &   $\cdots$    &  5$\pm$1\\
~3--1 S(3)           & 1.186    &   1.2$\pm$0.4 &  2.5$\pm$0.6 \\
~2--0 Q(1)           & 1.238    &  1.2$\pm$0.3  &  1.8$\pm$0.6 \\
~2--0 Q(2)           & 1.243    &  0.8$\pm$0.2  &  $\cdots$  \\
~2--0 Q(3)           & 1.247    &  1.1$\pm$0.3  &  1.5$\pm$0.4  \\
~2--0 Q(5)                   & 1.263        &  $\cdots$     &  2.4$\pm$0.8 \\
~2--0 Q(7)               & 1.287    &  $\cdots$     &  2.1$\pm$0.8$^a$   \\
~2--0 O(3)           & 1.335        &   $\cdots$    &  2.0$\pm$0.4\\
~1--0 S(9)           & 1.688    &  1.9$\pm$0.3  &  4.0$\pm$0.3  \\
~1--0 S(8)           & 1.715    &  1.3$\pm$0.2  & 2.8$\pm$0.4  \\
~1--0 S(7)           & 1.748    &  7.8$\pm$0.2  &  13.6$\pm$0.4  \\
~1--0 S(6)           & 1.788    &  7.7$\pm$0.8  &  11.2$\pm$0.4   \\
~1--0 S(3)           & 1.958    &  85$\pm$2     &   $\cdots$ \\
~2--1 S(4)           & 2.004    &  2.2$\pm$0.5  &   3.8$\pm$0.9  \\
~1--0 S(2)           & 2.034    &  19.7$\pm$0.4  &  23.9$\pm$0.9  \\
~3--2 S(5)           & 2.155    &  $\cdots$     &  1.2$\pm$0.4 \\
~2--1 S(3)           & 2.073    &  7.1$\pm$0.4  &  9.2$\pm$0.5   \\
~1--0 S(1)           & 2.122    & 48.2$\pm$0.7  &  58.1$\pm$0.8  \\
~2--1 S(2)           & 2.154    &  2.7$\pm$0.2  &  4$\pm$2$^a$  \\
~3--2 S(3)           & 2.201    &  1.5$\pm$0.4 &  2.2$\pm$0.7 \\
~1--0 S(0)           & 2.223    &  13.3$\pm$0.4 &  16$\pm$1  \\
~2--1 S(1)           & 2.248    &  8.4$\pm$0.4  &   11.7$\pm$0.7\\
~2--1 S(0)               & 2.356    &   $\cdots$    &  3$\pm$1 \\
~1--0 Q(1)           & 2.407    &   76$\pm$3    &  37$\pm$8 \\
~1--0 Q(2)           & 2.413    &   40$\pm$4    &  26$\pm$10$^a$ \\
~1--0 Q(3)           & 2.424    &  55$\pm$3     &  $\cdots$ \\
~1--0 Q(4)           & 2.437    &  39$\pm$4      &  34$\pm$7  \\
\hline\\[-5pt]
H$_2$ Lines          &          &   HH212 NK1   &  HH212 SK1   \\
\hline\\[-5pt]
~1--0 S(9)           & 1.688    &  2.1$\pm$0.3  &  2.6$\pm$0.5  \\
~1--0 S(8)           & 1.715    &  1.8$\pm$0.4  &  2.0$\pm$0.3  \\
~1--0 S(7)           & 1.748    &  8.3$\pm$0.4  &  10.3$\pm$0.4  \\
~1--0 S(6)           & 1.788    &  7.5$\pm$0.5  &  8.7$\pm$0.5   \\
~1--0 S(3)           & 1.958    &    90$\pm$3  &   32$\pm$4 \\
~2--1 S(4)           & 2.004    &  3.5$\pm$0.6  &  6.2$\pm$0.5  \\
~1--0 S(2)           & 2.034    &   27$\pm$1    &  27.1$\pm$0.9  \\
~2--1 S(3)           & 2.073    &  10.8$\pm$0.3  &  15.5$\pm$0.8   \\
~1--0 S(1)           & 2.122    &  62.4$\pm$0.5  &  64.3$\pm$0.6  \\
~2--1 S(2)           & 2.154    &  3.8$\pm$0.5  &  5.0$\pm$0.4  \\
~3--2 S(3)           & 2.201    &  2.6$\pm$0.4  &  3.2$\pm$0.5 \\
~1--0 S(0)           & 2.223    &  18$\pm$1     &  19.4$\pm$0.5 \\
~2--1 S(1)           & 2.248    &  13$\pm$1     &   18.3$\pm$0.5\\
~2--1 S(0)               & 2.356    & 2.3$\pm$0.8 &  4$\pm$1 \\
~3--2 S(1)               & 2.386    &  3$\pm$1     &  7$\pm$2 \\
~1--0 Q(1)           & 2.407    &   105$\pm$3    &  65$\pm$2 \\
~1--0 Q(2)           & 2.413    &    41$\pm$4    &  39$\pm$2 \\
~1--0 Q(3)           & 2.424    &  76$\pm$5     &  41$\pm$2 \\
~1--0 Q(4)           & 2.437    &  61$\pm$5     &  61$\pm$6  \\
~1--0 Q(5)           & 2.455    &  58$\pm$6     &   120$\pm$8  \\
~1--0 Q(6)           & 2.475    &  $\cdots$    &   44$\pm$10  \\
~1--0 Q(7)           & 2.500    &  $\cdots$     &   170$\pm$23  \\
\hline\\[-5pt]
~[\ion{Fe}{ii}]                  & 1.257       & 4.2$\pm$0.5    & 2.5$\pm$0.8\\
~[\ion{Fe}{ii}]                  & 1.534       & $\cdots$       & 1.0$\pm$0.3 \\
~[\ion{Fe}{ii}]                  & 1.600       & 0.8$\pm$0.2    & 0.8$\pm$0.2 \\
~[\ion{Fe}{ii}]                  & 1.644       & 7.7$\pm$0.4    & 6.0$\pm$0.4\\
~[\ion{Fe}{ii}]                  & 1.678       & $\cdots$       &  0.7$\pm$0.2 \\
\hline\\[-5pt]
\noindent
Notes: $^a$ Signal to noise ratio between 2 and 3.\\
\end{tabular}\\
\newpage
\end{table*}
\begin{table*}
\caption[]{ Observed lines in HH212 outflow
    \label{HH212sp2:tab}}
\begin{tabular}{cccccc}
\hline\\[-5pt]
Term&  $\lambda$($\mu$m) & \multicolumn{4}{c}{$F\pm\Delta~F$(10$^{-15}$erg\,cm$^{-2}$\,s$^{-1}$)}\\
\hline\\[-5pt]
H$_2$ Lines          &          &   HH212 NK2    &  HH212 NK4      &    HH212 NK7      & HH212 NB2 \\
\hline\\[-5pt]
~1--0 S(9)       & 1.688    &   $\cdots$  & $\cdots$     & $\cdots$   &  1.3$\pm$0.2  \\
~1--0 S(8)       & 1.715    & 0.5$\pm$0.1 &  $\cdots$    &  $\cdots$  & 0.9$\pm$0.2    \\
~1--0 S(7)       & 1.748    & 1.7$\pm$0.4 & 1.2$\pm$0.2  &  1.7$\pm$0.4 & 4.9$\pm$0.2        \\
~1--0 S(6)       & 1.788    & 1.7$\pm$0.5 &  $\cdots$  &  $\cdots$ & 4.9$\pm$0.3         \\
~1--0 S(3)       & 1.958    &  17$\pm$1   & 21$\pm$1     & 17$\pm$1 & 52$\pm$1 \\
~2--1 S(4)       & 2.004    & $\cdots$ & $\cdots$ & $\cdots$ & 1.5$\pm$0.3     \\
~1--0 S(2)       & 2.034    & 6.1$\pm$0.4 & 4.9$\pm$0.5  &  3.4$\pm$0.2 & 11.8$\pm$0.6       \\
~2--1 S(3)       & 2.073    & 2.2$\pm$0.4 & 2.2$\pm$0.7  &  1.2$\pm$0.2 & 4.2$\pm$0.4         \\
~1--0 S(1)       & 2.122    & 13.0$\pm$0.5 & 11.8$\pm$0.3 &  7.4$\pm$0.6 & 25.3$\pm$0.3    \\
~2--1 S(4)       & 2.154    & 0.9$\pm$0.2  &  $\cdots$   &  0.7$\pm$0.2    &  1.9$\pm$0.4 \\
~1--0 S(0)       & 2.223    & 3.6$\pm$0.3 & 4.1$\pm$0.4  &  3.2$\pm$0.9 & 7.1$\pm$0.3        \\
~2--1 S(1)       & 2.248    & 1.8$\pm$0.4 & 2.9$\pm$0.5  &  1.7$\pm$0.7$^a$ & 5.1$\pm$0.3      \\
~1--0 Q(1)       & 2.407    & 23$\pm$3   & 27$\pm$2      &  15$\pm$3 & 44$\pm$2        \\
~1--0 Q(2)       & 2.413    & 11$\pm$3   &  11$\pm$2     & $\cdots$ & 20$\pm$2         \\
~1--0 Q(3)       & 2.424    &  $\cdots$ & 20$\pm$2   & 14$\pm$3     & 36$\pm$2      \\
~1--0 Q(4)       & 2.437    & $\cdots$   &  15$\pm$3     & $\cdots$  & 19$\pm$3    \\
~1--0 Q(5)       & 2.455    &  $\cdots$ & $\cdots$ & $\cdots$    &  21$\pm$4    \\
\hline\\[-5pt]
\end{tabular}
\noindent
\\
\begin{tabular}{ccccccc}
\hline\\[-5pt]
 Line&  $\lambda$($\mu$m) & \multicolumn{5}{c}{$F\pm\Delta~F$(10$^{-15}$erg\,cm$^{-2}$\,s$^{-1}$)}\\
\hline\\[-5pt]
             &              &   HH212 SK2    &   HH212 SK4   &     HH212 SK5    &   HH212 SB1    &    HH212 SB2\\
\hline\\[-5pt]
~1--0 S(9)       & 1.688    & $\cdots$    &  $\cdots$  & $\cdots$   & $\cdots$ &   1.7$\pm$0.6$^a$ \\
~1--0 S(8)       & 1.715    & $\cdots$    &  $\cdots$  & $\cdots$   &   $\cdots$    &     1.1$\pm$0.3     \\
~1--0 S(7)       & 1.748    & $\cdots$    &  $\cdots$  & $\cdots$      & 5.3$\pm$0.5 &   6.3$\pm$0.4  \\
~1--0 S(6)       & 1.788    & $\cdots$    &  $\cdots$  &  $\cdots$    & 3.8$\pm$0.5 &   5.4$\pm$0.6   \\
~1--0 S(3)       & 1.958    & $\cdots$    &  $\cdots$   & $\cdots$ &    $\cdots$  &        9$\pm$4$^a$\\
~1--0 S(2)       & 2.034    & 2.7$\pm$0.4 & 2.1$\pm$0.3 & $\cdots$   & 11$\pm$2 &   15.5$\pm$0.8   \\
~2--1 S(3)       & 2.073    & 1.1$\pm$0.3   &  $\cdots$  & $\cdots$  &  3.6$\pm$0.6 &   5.4$\pm$0.7    \\
~1--0 S(1)       & 2.122    & 7.5$\pm$0.2 & 5.1$\pm$0.2 & 1.6$\pm$0.2  & 24.8$\pm$0.8 &   34.0$\pm$0.7   \\
~2--1 S(2)       & 2.154    & $\cdots$    &  $\cdots$  & $\cdots$   &  2$\pm$1$^a$ &    1.2$\pm$0.4       \\
~1--0 S(0)       & 2.223    & 3.1$\pm$0.7 & 1.5$\pm$0.2  & $\cdots$  & 7.9$\pm$0.8 &   9.7$\pm$0.5  \\
~2--1 S(1)       & 2.248    & 1.9$\pm$0.4 & $\cdots$  & $\cdots$     &  4.6$\pm$0.7 &   7.2$\pm$0.6 \\
~1--0 Q(1)       & 2.407    & $\cdots$   & $\cdots$   & $\cdots$      &   $\cdots$  &   $\cdots$   \\
~1--0 Q(2)       & 2.413    & $\cdots$   & $\cdots$   &  $\cdots$     &   $\cdots$ &   $\cdots$   \\
~1--0 Q(3)       & 2.424    & $\cdots$   & $\cdots$   &  $\cdots$    &   $\cdots$ &    $\cdots$     \\
~1--0 Q(4)       & 2.437    & $\cdots$   &  $\cdots$  &  $\cdots$     &     $\cdots$ &    $\cdots$   \\
~1--0 Q(5)       & 2.455    & $\cdots$   &  $\cdots$   &  $\cdots$     &   $\cdots$ &    $\cdots$     \\

\hline\\[-5pt]
\\
Notes: $^{a}$ S/N between 2 and 3.\\
\end{tabular}
\end{table*}


\begin{table*}
\caption[]{ Observed lines in NGC2264G outflow knots A2, A3, A7, C1 and C2
   \label{NGC2264Gsp:tab}}
\begin{tabular}{ccccccc}
\hline\\[-5pt]
Term&  $\lambda$($\mu$m) &
\multicolumn{4}{c}{$F\pm\Delta~F$(10$^{-15}$erg\,cm$^{-2}$\,s$^{-1}$)}\\
\hline\\[-5pt]
H$_2$ Lines          &          &   A2           &       A3         &    A7           &   C1        &   C2   \\
\hline\\[-5pt]
~1--0 S(9)           & 1.688        & 0.8$\pm$0.3$^a$ & $\cdots$        & $\cdots$        & $\cdots$     & $\cdots$    \\
~1--0 S(8)           & 1.715    & 0.8$\pm$0.3$^a$ & 6.7$\pm$0.2     & $\cdots$        & $\cdots$     & 0.6$\pm$0.2     \\
~1--0 S(7)           & 1.748    & 2.9$\pm$0.3     & 2.8$\pm$0.2     & 1.1$\pm$0.3$^a$ & 1.1$\pm$0.2      & 2.9$\pm$0.2  \\
~1--0 S(6)           & 1.788    & 2.2$\pm$0.3     & 2.0$\pm$0.2     & $\cdots$        & 0.8$\pm$0.3$^a$  & 2.0$\pm$0.2  \\
~1--0 S(3)           & 1.958    & 8.1$\pm$1       & $\cdots$        & 5.1$\pm$0.4     & $\cdots$     & 20$\pm$2      \\
~2--1 S(4)           & 2.004    & 9.9$\pm$0.3     & $\cdots$        & $\cdots$        & $\cdots$     &  $\cdots$        \\
~1--0 S(2)           & 2.034    & 5.0$\pm$0.2     & 5.4$\pm$0.3     & 1.7$\pm$0.4     & 1.1$\pm$0.3      & 5.2$\pm$0.2   \\
~2--1 S(3)           & 2.073        & 1.6$\pm$0.3     & 1.6$\pm$0.3     &  $\cdots$       & $\cdots$         & 0.9$\pm$0.2     \\
~1--0 S(1)           & 2.122    & 15.6$\pm$0.2    & 15.8$\pm$0.3    & 4.2$\pm$0.4     & 3.3$\pm$0.2      & 15.5$\pm$0.2    \\
~2--1 S(2)           & 2.154    & 0.8$\pm$0.2     & 1.0$\pm$0.2     &  $\cdots$       & 0.5$\pm$0.2$^a$  & 0.9$\pm$0.3    \\
~3--2 S(3)           & 2.201        &   $\cdots$      & 0.7$\pm$0.3$^a$ &  $\cdots$       & $\cdots$         &  $\cdots$       \\
~1--0 S(0)           & 2.223    & 4.0$\pm$0.2     & 4.1$\pm$0.3     & 1.1$\pm$0.3     & 0.9$\pm$0.2      & 4.0$\pm$0.3     \\
~2--1 S(1)           & 2.248    & 1.5$\pm$0.3     & 1.5$\pm$0.3     & 1.2$\pm$0.3     & $\cdots$         & 1.4$\pm$0.3     \\
~2--1 S(0)           & 2.355    & 0.8$\pm$0.3$^a$ &  $\cdots$       &  $\cdots$       & $\cdots$         & $\cdots$     \\
~3--2 S(1)               & 2.386    & 1.1$\pm$0.4$^a$ & 1.1$\pm$0.4$^a$ &  $\cdots$       & $\cdots$         & $\cdots$    \\
~1--0 Q(1)           & 2.407    & 3.8$\pm$0.3     & 4.5$\pm$0.5     &  $\cdots$       & $\cdots$         & 4.5$\pm$0.6       \\
~1--0 Q(2)           & 2.413    & 2.2$\pm$0.3     & 2.1$\pm$0.4     &  $\cdots$       & $\cdots$     & 2.3$\pm$0.5        \\
~1--0 Q(3)           & 2.424    & 3.2$\pm$0.3     & 2.9$\pm$0.4     &  $\cdots$       & $\cdots$     & 3.1$\pm$0.5        \\
~1--0 Q(4)           & 2.437    & 2.8$\pm$0.3     & 4.0$\pm$0.4     &  $\cdots$       & $\cdots$     & 3.8$\pm$0.5     \\
~1--0 Q(5)           & 2.455    & 7.5$\pm$0.3     & 8.3$\pm$0.5     &  $\cdots$       & $\cdots$     &  $\cdots$     \\

\hline\\[-5pt]
\\
Notes: $^{a}$S/N between 2 and 3.\\
\end{tabular}\\
\end{table*}

\begin{table*}
\caption[]{Observed lines in VLA1623 HH313 A
\label{HH313A:tab}}
\begin{tabular}{ccc}
\hline\\[-5pt]
Term&  $\lambda$($\mu$m) & $F\pm\Delta~F$(10$^{-15}$erg\,cm$^{-2}$\,s$^{-1}$)\\
\hline\\[-5pt]
H$_2$ Lines          &          &   HH313 A     \\
\hline\\[-5pt]
~2--0 S(9)                       & 1.053        & 1.7$\pm$0.6$^a$     \\
~2--0 S(7)               & 1.064        & 1.3$\pm$0.6$^a$     \\
~2--0 S(6)               & 1.073        & 2.2$\pm$0.7          \\
~2--0 S(5)               & 1.085        & 5.6$\pm$1.5          \\
~2--0 S(4)           & 1.100    & 3.0$\pm$1.0            \\
~2--0 S(3)           & 1.117    & 3.1$\pm$0.8           \\
~3--1 S(9-10-11)                 & 1.120-1.121  & 4.3$\pm$0.8           \\
~3--1 S(7)           & 1.130    & 2.7$\pm$0.5           \\
~2--0 S(2)+3--1 S(6)         & 1.138-1.140  & 5.0$\pm$1.0       \\
~3--1 S(5)           & 1.152    & 3.2$\pm$0.7           \\
~2--0 S(1)           & 1.162    & 3.5$\pm$0.5             \\
~3--1 S(3)           & 1.186    & 2.9$\pm$0.5           \\
~3--1 S(1)           & 1.233    & 2.1$\pm$0.4           \\
~2--0 Q(1)           & 1.238    & 3.3$\pm$0.5        \\
~2--0 Q(2)           & 1.242        & 1.4$\pm$0.5$^a$         \\
~2--0 Q(3)           & 1.247    & 3.0$\pm$0.4           \\
~2--0 Q(4)           & 1.254    & 1.2$\pm$0.4            \\
~2--0 Q(5)+3--1 S(0)             & 1.261-1.263  & 3.3$\pm$0.7             \\
~2--0 Q(7)               & 1.287    & 2.5$\pm$0.3            \\
~4--2 S(1)           & 1.311    & 1.9$\pm$0.6           \\
~2--0 O(3)               & 1.335        & 3.0$\pm$0.3            \\
~3--1 Q(5)                   & 1.342        & 1.9$\pm$0.4           \\
~5--3 S(3)           & 1.347    & 1.2$\pm$0.4       \\
~1--0 S(11)              & 1.650    & 2.7$\pm$0.9        \\
~1--0 S(10)              & 1.666    & 5.6$\pm$0.6        \\
~1--0 S(9)           & 1.688    &  13$\pm$1            \\
~1--0 S(8)           & 1.715    &  9$\pm$1        \\
~1--0 S(7)           & 1.748    & 37$\pm$1            \\
~1--0 S(6)           & 1.788    & 32$\pm$1            \\
~1--0 S(3)           & 1.958    & 38$\pm$5               \\
~2--1 S(4)           & 2.004    & 17$\pm$2             \\
~1--0 S(2)           & 2.034    & 76$\pm$1            \\
~3--2 S(5)           & 2.066    & 3.8$\pm$0.6         \\
~2--1 S(3)           & 2.073        & 22.6$\pm$0.5             \\
~1--0 S(1)           & 2.122    & 215$\pm$2       \\
~3--2 S(4)           & 2.127    &  5$\pm$1        \\
~2--1 S(2)           & 2.154    & 12$\pm$1        \\
~3--2 S(3)           & 2.201    & 5.2$\pm$0.9      \\
~1--0 S(0)           & 2.223    & 41.0$\pm$0.8            \\
~2--1 S(1)           & 2.248    & 21.4$\pm$0.8            \\
~2--1 S(0)           & 2.355    & 8.5$\pm$1.2         \\
~3--2 S(1)               & 2.386    & 7$\pm$1           \\
~1--0 Q(1)           & 2.407    & 193$\pm$5          \\
~1--0 Q(2)           & 2.413    &  97$\pm$5           \\
~1--0 Q(3)           & 2.424    &  213$\pm$5              \\
~1--0 Q(4)           & 2.437    &  76$\pm$5            \\
~1--0 Q(5)           & 2.455    &  116$\pm$8            \\

\hline\\[-5pt]
  [{\ion{Fe}{II}}] lines &  $\lambda$($\mu$m) & HH313 A \\
\hline\\[-5pt]
~$a^4\!D_{7/2}-a^6\!D_{9/2}$ & 1.257 & 5.0$\pm$0.4\\
~$a^4\!D_{7/2}-a^6\!D_{7/2}$ & 1.321 & 3.0$\pm$0.6 \\
~$a^4\!D_{5/2}-a^4\!F_{9/2}$ & 1.534 & 5$\pm$2$^a$ \\
~$a^4\!D_{7/2}-a^4\!F_{9/2}$ & 1.644 & 15$\pm$1 \\
~$a^4\!D_{5/2}-a^4\!F_{7/2}$ & 1.678 & 1.0$\pm$0.5$^a$ \\
\hline\\[-5pt]
Other ionic lines               & &    \\
\hline\\[-5pt]
\hline\\[-5pt]
~[\ion{C}{I}]\,$^1\!D_{2}-^3\!P_{2}$ & 0.985 &  6$\pm$1 \\
\hline\\[-5pt]
\\
Notes: $^a$ S/N ratio between 2 and 3. \\
\end{tabular}\\
\end{table*}

\begin{table*}
\caption[]{Observed lines in VLA1623 knots HH313 B, GSWC2003-13a, 13b, 14a, 14c+d, 14e
\label{VLA1623sp1:tab}}
\begin{tabular}{cccccccc}
\hline\\[-5pt]
Term&  $\lambda$($\mu$m) &
\multicolumn{6}{c}{$F\pm\Delta~F$(10$^{-15}$erg\,cm$^{-2}$\,s$^{-1}$)}\\
\hline\\[-5pt]
H$_2$ Lines     &          &  HH313 B       &  13a            &  13b        & 14a        & 14c+14d     &   14e       \\
\hline\\[-5pt]
~1--0 S(9)      & 1.688    & 1.8$\pm$0.5     & $\cdots$       & $\cdots$    & $\cdots$     & $\cdots$           & 1.3$\pm$0.4        \\
~1--0 S(8)      & 1.715    & 4.5$\pm$0.7     & $\cdots$       & $\cdots$    & $\cdots$     & $\cdots$           & 1.1$\pm$0.4$^a$     \\
~1--0 S(7)      & 1.748    & 20.0$\pm$0.7    & 4.9$\pm$0.5    & $\cdots$    & 2.5$\pm$0.5    & 3.9$\pm$0.5      & 5.6$\pm$0.5        \\
~1--0 S(6)      & 1.788    & 13.3$\pm$0.8    & 3.7$\pm$0.6    & $\cdots$    & 1.9$\pm$0.5    & 2.4$\pm$0.9      & 3.3$\pm$0.7        \\
~1--0 S(3)      & 1.958    & 106$\pm$5       & 34$\pm$2       & $\cdots$    & 18$\pm$1     & 30$\pm$2           & 20$\pm$2           \\
~1--0 S(2)      & 2.034    &  45$\pm$2       & 7.9$\pm$0.8    & $\cdots$    & 4.4$\pm$0.7    & 7.0$\pm$1.0      & 10.7$\pm$0.4      \\
~3--2 S(5)      & 2.066    & 1.6$\pm$0.5     &  $\cdots$      & $\cdots$    &  $\cdots$    & $\cdots$           &  $\cdots$       \\
~2--1 S(3)      & 2.073    & 12.7$\pm$0.6    & 2.8$\pm$0.4    & $\cdots$    & 1.3$\pm$0.6$^a$ & 1.7$\pm$0.6$^a$ & 4.6$\pm$0.4       \\
~1--0 S(1)      & 2.122    & 128.0$\pm$0.5   & 19.5$\pm$0.4   & 3.0$\pm$0.5 & 12.1$\pm$0.4   & 18.2$\pm$0.6     & 30.2$\pm$0.3      \\
~2--1 S(2)      & 2.154    & 5.1$\pm$0.5     &1.4$\pm$0.5$^a$ & $\cdots$    & $\cdots$     &  $\cdots$        & 2.6$\pm$0.3     \\
~3--2 S(3)      & 2.201    & 3.5$\pm$0.4     & $\cdots$       & $\cdots$    & $\cdots$       & $\cdots$     &  1.2$\pm$0.3         \\
~1--0 S(0)      & 2.223    & 30.6$\pm$0.6    & 4.0$\pm$0.5    & $\cdots$    & 3.6$\pm$0.5    & 4.1$\pm$0.6  & 7.9$\pm$0.3         \\
~2--1 S(1)      & 2.248    & 14.8$\pm$0.6    & 3.3$\pm$0.8    & $\cdots$    & $\cdots$       & 2.3$\pm$0.8$^a$  & 5.6$\pm$0.3         \\
~2--1 S(0)      & 2.355    & 3.5$\pm$1.1     & $\cdots$       & $\cdots$    & $\cdots$       & $\cdots$     & 1.9$\pm$0.4         \\
~3--2 S(1)          & 2.386    & 3.6$\pm$1.1     & $\cdots$       & $\cdots$    & $\cdots$       & $\cdots$     &  $\cdots$           \\
~1--0 Q(1)      & 2.407    & 139$\pm$5       &  21$\pm$4      & $\cdots$    &  12$\pm$4      & 19$\pm$4     & 33$\pm$2        \\
~1--0 Q(2)      & 2.413    &  57$\pm$5       &  $\cdots$      & $\cdots$    &  $\cdots$      & $\cdots$     & 12$\pm$2         \\
~1--0 Q(3)      & 2.424    & 140$\pm$5       &  19$\pm$4      & $\cdots$    &  12$\pm$4      & 18$\pm$4     & 33$\pm$2         \\
~1--0 Q(4)      & 2.437    &  45$\pm$5       & $\cdots$       & $\cdots$    & $\cdots$       &  $\cdots$    & 10$\pm$2           \\
~1--0 Q(5)      & 2.455    &  41$\pm$5       & $\cdots$       & $\cdots$    & $\cdots$       &  $\cdots$    & 14$\pm$2           \\
\hline\\[-5pt]
\end{tabular}
\\
Notes: $^{a}$ S/N between 2 and 3.\\
\end{table*}
\begin{table*}
\caption[]{Observed lines in VLA1623 knots GSWC2003-14f, 14g, 14h, 18, 20b, 20c, 20d
\label{VLA1623sp2:tab}}
\begin{tabular}{ccccccccc}
\hline\\[-5pt]
H$_2$ Lines     &          &  14f       &  14g        &  14h        &  18           & 20b         & 20c           &   20d   \\
\hline\\[-5pt]

~1--0 S(7)      & 1.748    & 2.2$\pm$0.6  & 10.3$\pm$0.9    & 4.0$\pm$1.0     & $\cdots$       & $\cdots$       & 7.4$\pm$0.5     & 4.8$\pm$0.5       \\
~1--0 S(6)      & 1.788    & $\cdots$     & 9.0$\pm$0.8     & 4.6$\pm$1.5     & 1.7$\pm$0.6$^a$& $\cdots$       & 6.1$\pm$0.6     & 4.3$\pm$0.6       \\
~1--0 S(3)      & 1.958    & $\cdots$     & 19.3$\pm$3      & $\cdots$        & 10$\pm$2       & $\cdots$       & 8$\pm$2         & 27$\pm$4        \\
~1--0 S(2)      & 2.034    & 2.7$\pm$0.7  & 13.3$\pm$0.6    & 5.0$\pm$1.0     & 3.8$\pm$0.5    &1.4$\pm$0.5$^a$ & 5.7$\pm$0.6     & 9.6$\pm$0.6       \\
~2--1 S(3)      & 2.073    & $\cdots$     & 6.8$\pm$0.9     & $\cdots$        & 2.0$\pm$0.6    & $\cdots$       & 1.7$\pm$0.6$^a$ & 5.7$\pm$0.6       \\
~1--0 S(1)      & 2.122    & 10.3$\pm$0.3 & 43.2$\pm$0.5    & 17$\pm$1        & 11.9$\pm$0.8   & 3.2$\pm$0.4    & 13.0$\pm$0.8    & 34.5$\pm$0.8        \\
~2--1 S(2)      & 2.154    & $\cdots$     & 4.1$\pm$0.7     & 1.7$\pm$0.7$^a$ & $\cdots$       & $\cdots$       &  $\cdots$   & 2.1$\pm$0.5      \\
~3--2 S(3)      & 2.201    & $\cdots$     & 2.2$\pm$0.9$^a$ & $\cdots$        & $\cdots$       & $\cdots$       & $\cdots$    & 1.4$\pm$0.4  \\
~1--0 S(0)      & 2.223    & 2.2$\pm$0.4  & 10.3$\pm$0.4    & 5.0$\pm$1.0     & 4.4$\pm$0.8    & $\cdots$       & 3.8$\pm$0.5     & 9.0$\pm$0.7    \\
~2--1 S(1)      & 2.248    & 1.6$\pm$0.5  & 8.9$\pm$0.6     & 4$\pm$1         & 2.4$\pm$0.8    & 1.6$\pm$0.5    & $\cdots$        & 4.1$\pm$0.7    \\
~1--0 Q(1)      & 2.407    & 13$\pm$2     &  48$\pm$2       & 4$\pm$1         & 17$\pm$2       & $\cdots$       & 17$\pm$4    & 37$\pm$4          \\
~1--0 Q(2)      & 2.413    & $\cdots$     &  24$\pm$2       & $\cdots$        & 4$\pm$2$^a$    & $\cdots$       &  6$\pm$2    & 15$\pm$4          \\
~1--0 Q(3)      & 2.424    & 13$\pm$2     &  55$\pm$2       & $\cdots$        & 18$\pm$2       & $\cdots$       & 19$\pm$3        & 41$\pm$4      \\
~1--0 Q(4)      & 2.437    & $\cdots$     &  25$\pm$5       & $\cdots$        & 6$\pm$3        & $\cdots$       &  $\cdots$   & 10$\pm$4$^a$          \\
~1--0 Q(5)      & 2.455    & $\cdots$     &  32$\pm$5       & $\cdots$        & 15$\pm$4       & $\cdots$       &  13$\pm$5   & 27$\pm$5        \\
\hline\\[-5pt]
\end{tabular}
\\
Notes: $^{a}$S/N between 2 and 3.\\
\end{table*}

\begin{table*}
\caption[]{Observed lines in VLA1623 knots GSWC2003-20e, 20f
\label{VLA1623sp3:tab}}
\begin{tabular}{cccc}
\hline\\[-5pt]
H$_2$ Lines     &          & 20e        &  20f    \\
\hline\\[-5pt]
~1--0 S(9)      & 1.688  & $\cdots$     & 1.7$\pm$0.7$^a$ \\
~1--0 S(8)      & 1.715  & $\cdots$     & $\cdots$    \\
~1--0 S(7)      & 1.748  & 1.3$\pm$0.5$^a$  & 4.8$\pm$0.7   \\
~1--0 S(6)      & 1.788  & 1.3$\pm$0.5$^a$  & 4.1$\pm$0.8   \\
~1--0 S(3)      & 1.958  & 5$\pm$2$^a$      & 10$\pm$2       \\
~1--0 S(2)      & 2.034  & 2.6$\pm$0.5      & 16.7$\pm$0.6        \\
~3--2 S(5)      & 2.066  &  $\cdots$        & $\cdots$      \\
~2--1 S(3)      & 2.073  & 0.8$\pm$0.3$^a$  & 5.1$\pm$0.6    \\
~1--0 S(1)      & 2.122  & 7.7$\pm$0.4      & 46.0$\pm$0.5    \\
~2--1 S(2)      & 2.154  & $\cdots$         & 1.9$\pm$0.4      \\
~3--2 S(3)      & 2.201  & $\cdots$     & 1.0$\pm$0.4$^a$    \\
~1--0 S(0)      & 2.223  & 1.9$\pm$0.5      & 13.6$\pm$0.6   \\
~2--1 S(1)      & 2.248  & $\cdots$         & 5.4$\pm$0.6    \\
~2--1 S(0)      & 2.355  & $\cdots$     &   $\cdots$     \\
~3--2 S(1)          & 2.386  & $\cdots$     &   $\cdots$     \\
~1--0 Q(1)      & 2.407  &  8$\pm$2     & 59$\pm$5   \\
~1--0 Q(2)      & 2.413  &  4$\pm$2$^a$     &  23$\pm$5      \\
~1--0 Q(3)      & 2.424  & 10$\pm$2     & 58$\pm$5  \\
~1--0 Q(4)      & 2.437  &  $\cdots$        &  20$\pm$5       \\
~1--0 Q(5)      & 2.455  &  7$\pm$2     &  32$\pm$5       \\
\hline\\[-5pt]
\end{tabular}
\\
Notes: $^{a}$S/N between 2 and 3.\\
\end{table*}
\begin{table*}
\caption[]{ [{\ion{Fe}{II}}] observed lines in knots HH313 B, GSWC2003-14g, 14h 
\label{VLAFe:tab}}
\begin{tabular}{ccccc}
\hline\\[-5pt]
 Line&  $\lambda$($\mu$m) & \multicolumn{3}{c}{$F \pm \Delta~F$(10$^{-15}$ erg\,cm$^{-2}$\,s$^{-1}$)}\\[+5pt]
\hline\\[-5pt]
  [{\ion{Fe}{II}}] lines     &       &   HH313 B    &   14g           &     14h      \\
\hline\\[-5pt]
~$a^4\!D_{7/2}-a^6\!D_{9/2}$ & 1.257 &   $\cdots$   & 1.4$\pm$0.6$^a$ & 1.5$\pm$0.5     \\
~$a^4\!D_{7/2}-a^4\!F_{9/2}$ & 1.644 & 1.8$\pm$0.5  &  6.0$\pm$0.7    & 7.2$\pm$0.5     \\
\hline\\[-5pt]
Notes:$^{a}$S/N between 2 and 3.\\
\end{tabular}
\end{table*}


\begin{table*}
\caption[]{ Observed lines in IRAS18273+0113 outflow knots S6 and S7
\label{IRAS18273sp:tab}}
\begin{tabular}{cccc}
\hline\\[-5pt]
Term&  $\lambda$($\mu$m) &
\multicolumn{2}{c}{$F\pm\Delta~F$(10$^{-15}$erg\,cm$^{-2}$\,s$^{-1}$)}\\
\hline\\[-5pt]
H$_2$ Lines      &              &  S6            &    S7             \\
~1--0 S(7)       & 1.748    & 3.3$\pm$0.7    & $\cdots$        \\
~1--0 S(6)       & 1.788    & 3.4$\pm$0.8    &  2.8$\pm$0.8         \\
~1--0 S(3)       & 1.958    & 37$\pm$2       & 17$\pm$1  \\
~1--0 S(2)       & 2.034    & 14.9$\pm$1.2   &  7$\pm$1    \\
~2--1 S(3)       & 2.073    &  3.2$\pm$0.6   &  2.9$\pm$0.5          \\
~1--0 S(1)       & 2.122    & 41.1$\pm$0.5   &  14.8$\pm$0.3    \\
~2--1 S(4)       & 2.154    & 1.3$\pm$0.3    &  $\cdots$  \\
~1--0 S(0)       & 2.223    & 11.1$\pm$0.7   &  5.2$\pm$0.8   \\
~2--1 S(1)       & 2.248    & 5.3$\pm$0.4    &  2.8$\pm$0.6       \\
~2--1 S(0)       & 2.355    & 1.1$\pm$0.3    & $\cdots$  \\
~1--0 Q(1)       & 2.407    & 37$\pm$2       &  14$\pm$1       \\
~1--0 Q(2)       & 2.413    & 15$\pm$2       & 5$\pm$1        \\
~1--0 Q(3)       & 2.424    & 35$\pm$2       & 11$\pm$1        \\
~1--0 Q(4)       & 2.437    &  9$\pm$1   & $\cdots$      \\
~1--0 Q(5)       & 2.455    &  9$\pm$1    & $\cdots$    \\
\hline\\
\end{tabular}
\\
Notes: $^{a}$S/N between 2 and 3.\\
\end{table*}


\begin{table*}
\caption[]{ Observed lines in L1157 knots A1, A2, C1, C2 and C4
\label{L1157sp1:tab}}
\begin{tabular}{ccccccc}
\hline\\[-5pt]
Term&  $\lambda$($\mu$m) & \multicolumn{5}{c}{$F\pm\Delta~F$(10$^{-15}$erg\,cm$^{-2}$\,s$^{-1}$)}\\
\hline\\[-5pt]
H$_2$ Lines          &          &   A1           &       A2       &   C1            & C2        &   C4   \\
\hline\\[-5pt]
~2--0 S(4)           & 1.100    & 2.4$\pm$0.4    & 2.5$\pm$0.8    & $\cdots$       & $\cdots$        & $\cdots$        \\
~2--0 S(3)           & 1.117    & 7.8$\pm$0.8    & 6.0$\pm$2.0    & $\cdots$       & $\cdots$        & $\cdots$       \\
~3--1 S(7)           & 1.130    & 2.5$\pm$0.8    &  $\cdots$      & $\cdots$       & $\cdots$        & $\cdots$       \\
~2--0 S(2)+3--1 S(6)         & 1.138-1.140  & 4.4$\pm$2.0$^a$&  $\cdots$      & $\cdots$       & $\cdots$        & $\cdots$       \\
~3--1 S(5)           & 1.152    & 5.0$\pm$1.0    & 2.9$\pm$0.9    & 3.3$\pm$1.1    & $\cdots$        & $\cdots$       \\
~2--0 S(1)           & 1.162    & 4.9$\pm$1.5    & 3.4$\pm$0.5    & 6.4$\pm$2.0    & 5.3$\pm$1.5     & 1.7$\pm$0.5      \\
~3--1 S(3)           & 1.186    & 3.6$\pm$0.6    & 2.3$\pm$0.5    & 2.3$\pm$1.0$^a$& $\cdots$        & $\cdots$       \\
~4--2 S(9)           & 1.196    & 1.3$\pm$0.4    & 1.0$\pm$0.5$^a$&$\cdots$        & $\cdots$        & $\cdots$         \\
~4--2 S(7)+3--1 S(2)         & 1.205-1.207  & 1.2$\pm$0.5$^a$& 2.0$\pm$0.6    & $\cdots$       & $\cdots$        & $\cdots$       \\
~4--2 S(5)           & 1.226    & 1.8$\pm$0.5    & 1.8$\pm$0.5    & 2.1$\pm$0.8$^a$& $\cdots$        & $\cdots$       \\
~3--1 S(1)           & 1.233    & 2.6$\pm$0.6    & 2.3$\pm$0.5    & 1.6$\pm$0.8$^a$& $\cdots$        & $\cdots$       \\
~2--0 Q(1)           & 1.238    & 4.0$\pm$0.6    & 2.1$\pm$0.4    & 4$\pm$1        & 4.8$\pm$1.5     & 1.2$\pm$0.4     \\
~2--0 Q(2)           & 1.242        & 2.8$\pm$0.5    & 1.3$\pm$0.4    & $\cdots$       & $\cdots$        & $\cdots$        \\
~2--0 Q(3)           & 1.247    & 3.2$\pm$0.4    & 2.3$\pm$0.4    & 5$\pm$1        & 3$\pm$1         & 1$\pm$0.4$^a$          \\
~2--0 Q(4)           & 1.254    & 2.1$\pm$0.5    & 1.0$\pm$0.4$^a$ &1.6$\pm$0.6$^a$& $\cdots$        & $\cdots$        \\
~4--2 S(3)+2--0 Q(5)+3--1 S(0)   & 1.261-1.263  & 3.8$\pm$0.5    & 3.8$\pm$0.5    & 3.2$\pm$1.0    & $\cdots$        & $\cdots$         \\
~4--2 S(2)+2--0 Q(7)         & 1.285-1.287  & 3.4$\pm$0.7    & 1.5$\pm$0.5    & 3.3$\pm$1.0    & $\cdots$        & $\cdots$        \\
~4--2 S(1)           & 1.311    & 1.4$\pm$0.3    & 2.5$\pm$0.8    & 1.9$\pm$0.9$^a$& $\cdots$        & $\cdots$       \\
~3--1 Q(1)               & 1.314    & 2.4$\pm$0.4    &  $\cdots$      & $\cdots$       & $\cdots$        & $\cdots$       \\
~2--0 Q(9)           & 1.319    & 2.8$\pm$0.8    & 1.8$\pm$0.6    & 2.6$\pm$0.9$^a$& $\cdots$        & $\cdots$       \\
~3--1 Q(3)           & 1.324    & 1.6$\pm$0.8$^a$& $\cdots$       & $\cdots$       & $\cdots$        & $\cdots$       \\
~2--0 O(3)               & 1.335        & 2.7$\pm$0.7    & 2.3$\pm$0.6    & 2.7$\pm$0.9    & $\cdots$        & $\cdots$        \\
~3--1 Q(5)+4--2 S(0)         & 1.342-1.342  & 2.3$\pm$0.7    & 1.7$\pm$0.7$^a$&1.5$\pm$0.9$^a$ & $\cdots$        & $\cdots$       \\
~5--3 S(3)           & 1.347    & 1.4$\pm$0.7$^a$&   2$\pm$1$^a$  &$\cdots$        & $\cdots$        & $\cdots$       \\
~1--0 S(11)              & 1.650    & 1.1$\pm$0.4$^a$& 1.2$\pm$0.4    & $\cdots$       & $\cdots$        & $\cdots$        \\
~1--0 S(9)           & 1.688    & 4.3$\pm$0.8    & 4.3$\pm$0.5    & 4.4$\pm$0.8    & 3.5$\pm$0.8     & 1.4$\pm$0.4     \\
~1--0 S(8)           & 1.715    & 4.1$\pm$0.8    & 3.4$\pm$0.8    & 4.2$\pm$0.9    & 5.1$\pm$0.7     & 1.0$\pm$0.4$^a$          \\
~1--0 S(7)           & 1.748    & 16.7$\pm$0.8   & 14.2$\pm$0.6   &13.3$\pm$0.9    & 15.6$\pm$0.5    & 4.3$\pm$0.5      \\
~1--0 S(6)           & 1.788    & 13.0$\pm$0.8   & 10.3$\pm$0.8   & 9.6$\pm$0.7    & 10.6$\pm$0.5    & 3.2$\pm$0.7      \\
~1--0 S(5)           & 1.835$^b$    & 106$\pm$20     & 98$\pm$23      & $\cdots$       & $\cdots$        & $\cdots$             \\
~1--0 S(4)           & 1.891$^b$    & 105$\pm$20     & 88$\pm$16      & $\cdots$       & $\cdots$        & $\cdots$         \\
~1--0 S(3)           & 1.958    & 136$\pm$5      & 132$\pm$16     & 73$\pm$5       & 87$\pm$5        & 20$\pm$5         \\
~2--1 S(4)           & 2.004    & 4.6$\pm$0.7    & 2.7$\pm$0.7    & $\cdots$       &  $\cdots$       &  $\cdots$        \\
~1--0 S(2)           & 2.034    & 21.0$\pm$0.4   & 15.9$\pm$0.9   & 17.0$\pm$0.6   & 19.7$\pm$0.8    & 6.0$\pm$0.8      \\
~3--2 S(5)           & 2.066    & 1.6$\pm$0.3    & 1.9$\pm$0.4    &  $\cdots$      & $\cdots$        &  $\cdots$        \\
~2--1 S(3)           & 2.073        & 6.2$\pm$0.4    & 6.2$\pm$0.5    &  6.0$\pm$0.5   & 7.8$\pm$0.8     & 1.8$\pm$0.5      \\
~1--0 S(1)           & 2.122    & 52.6$\pm$0.6   & 40.2$\pm$0.5   & 44.6$\pm$0.4   & 53.6$\pm$0.5    & 13.8$\pm$0.3     \\
~2--1 S(2)           & 2.154    & 2.4$\pm$0.4    & 2.1$\pm$0.3    &  2.0$\pm$0.4   &  2.2$\pm$0.5    &  0.9$\pm$0.3     \\
~3--2 S(3)           & 2.201    & 1.3$\pm$0.3    & 2.5$\pm$0.3    & 1.3$\pm$0.6$^a$& 1.9$\pm$0.3     &  $\cdots$     \\
~1--0 S(0)           & 2.223    & 12.1$\pm$0.4   & 9.6$\pm$0.5    & 10.3$\pm$0.5   & 12.4$\pm$0.3    & 3.4$\pm$0.3      \\
~2--1 S(1)           & 2.248    &  6.6$\pm$0.5   & 6.7$\pm$0.5    &  5.1$\pm$0.5   & 6.3$\pm$0.6     & 1.9$\pm$0.4      \\
~2--1 S(0)           & 2.355    & 1.6$\pm$0.8$^a$& 2.5$\pm$0.8    &   $\cdots$     & 2.1$\pm$0.7     &  $\cdots$        \\
~3--2 S(1)               & 2.386    & 1.4$\pm$0.6$^a$& 1.8$\pm$0.6    &   $\cdots$     & 2.3$\pm$0.8$^a$ &  $\cdots$    \\
~1--0 Q(1)           & 2.407    &   58$\pm$5     &  41$\pm$5      &  37$\pm$5      & 49$\pm$5        & 13$\pm$3            \\
~1--0 Q(2)           & 2.413    &   24$\pm$5     &  14$\pm$5      &  14$\pm$5      & 21$\pm$5        & $\cdots$         \\
~1--0 Q(3)           & 2.424    &   53$\pm$5     &  39$\pm$5      &  30$\pm$5      & 39$\pm$5        & 11$\pm$3         \\
~1--0 Q(4)           & 2.437    &   27$\pm$5     &  19$\pm$5      &  14$\pm$5      &  18$\pm$5       &  $\cdots$      \\
~1--0 Q(5)           & 2.455    &   27$\pm$5     &  19$\pm$5      &  10$\pm$5$^a$  &  $\cdots$       &  $\cdots$      \\
\hline\\[-5pt]
  [{\ion{Fe}{II}}] lines         &              &                &                &                &                  &     \\
\hline\\[-5pt]
~$a^4\!D_{7/2}-a^6\!D_{9/2}$ & 1.257 & 0.5$\pm$0.2$^a$   & 3.6$\pm$0.5 &   $\cdots$  &   $\cdots$ &   $\cdots$      \\
~$a^4\!D_{7/2}-a^4\!F_{9/2}$ & 1.644 & 0.6$\pm$0.3$^a$  & 3.8$\pm$0.6 &   $\cdots$  &   $\cdots$  &   $\cdots$      \\
\hline\\[-5pt]
\end{tabular}
\\
Notes: $^{a}$S/N between 2 and 3.\\
\end{table*}


\clearpage
\begin{table*}
\caption[]{ Observed lines in L1157 knots A4, A8 and D3
\label{L1157sp2:tab}}
\begin{tabular}{ccccc}
\hline\\[-5pt]
Term&  $\lambda$($\mu$m) &
\multicolumn{3}{c}{$F\pm\Delta~F$(10$^{-15}$erg\,cm$^{-2}$\,s$^{-1}$)}\\
\hline\\[-5pt]
H$_2$ Lines          &          &   A4           &       A8     &     D3              \\
\hline\\[-5pt]
~1--0 S(7)           & 1.748    & 1.5$\pm$0.3   & 1.4$\pm$0.4   & 3.1$\pm$0.9        \\
~1--0 S(6)           & 1.788    & 1.3$\pm$0.3   &  $\cdots$     & 3.3$\pm$1.0        \\
~1--0 S(3)           & 1.958    & 28$\pm$5      & 10$\pm$3      & 30$\pm$5           \\
~1--0 S(2)           & 2.034    & 3.8$\pm$0.4   & 1.3$\pm$0.4   & 5.8$\pm$0.7        \\
~2--1 S(3)           & 2.073        & 0.8$\pm$0.3$^a$ & $\cdots$    & 2.5$\pm$0.7        \\
~1--0 S(1)           & 2.122    & 7.7$\pm$0.3   & 3.2$\pm$0.4   & 12.0$\pm$0.4       \\
~2--1 S(2)           & 2.154    & 0.6$\pm$0.3$^a$& $\cdots$     & 0.6$\pm$0.3$^a$        \\
~1--0 S(0)           & 2.223    & 2.5$\pm$0.4  & 0.9$\pm$0.3    & 2.7$\pm$0.5        \\
~2--1 S(1)           & 2.248    & 0.8$\pm$0.3$^a$  & $\cdots$   & $\cdots$           \\
~1--0 Q(1)           & 2.407    &  9$\pm$3     &  $\cdots$      &  13$\pm$4         \\
~1--0 Q(3)           & 2.424    &  8$\pm$3     &  $\cdots$      &  8$\pm$4$^a$           \\

\hline\\[-5pt]
\end{tabular}
\\
Notes: $^{a}$S/N between 2 and 3.\\
\end{table*}


\begin{table*}
\caption[]{ Observed lines in IC1396N outflow knots A and B
\label{IC1396sp:tab}}
\begin{tabular}{cccc}
\hline\\[-5pt]
Term&  $\lambda$($\mu$m) &
\multicolumn{2}{c}{$F\pm\Delta~F$(10$^{-15}$erg\,cm$^{-2}$\,s$^{-1}$)}\\
\hline\\[-5pt]
H$_2$ Lines          &           &    A            &    B             \\
~1--0 S(9)           & 1.688     & 1.5$\pm$0.3     &  $\cdots$         \\
~1--0 S(7)           & 1.748     & 3.2$\pm$0.3     & 1.6$\pm$0.3        \\
~1--0 S(6)           & 1.788     & 3.2$\pm$0.4     & 1.4$\pm$0.4        \\
~1--0 S(3)           & 1.958     & 33$\pm$3        & 13$\pm$1       \\
~2--1 S(4)           & 2.004     & 1.2$\pm$0.4     &  $\cdots$      \\
~1--0 S(2)           & 2.034     & 6.5$\pm$0.4     & 3.6$\pm$0.4        \\
~2--1 S(3)           & 2.073     & 1.0$\pm$0.3     & $\cdots$   \\
~1--0 S(1)           & 2.122     & 17.3$\pm$0.3    & 9.1$\pm$0.4        \\
~2--1 S(2)           & 2.154     & 1.0$\pm$0.3     &  $\cdots$      \\
~3--2 S(3)           & 2.201     & 0.7$\pm$0.3$^a$ &  $\cdots$       \\
~1--0 S(0)           & 2.223     & 4.2$\pm$0.3     &  2.4$\pm$0.3       \\
~1--0 Q(1)           & 2.407     & 19$\pm$2       &   11$\pm$2         \\
~1--0 Q(2)           & 2.413     & 8$\pm$2       &   $\cdots$       \\
~1--0 Q(3)           & 2.424     & 18$\pm$2       &   9$\pm$2       \\

\hline\\[-5pt]
Atomic and Ionic lines           &      &                 &          \\
\hline\\[-5pt]
~[{\ion{Fe}{ii}}]\,$a^4\!D_{7/2}-a^4\!F_{9/2}$  & 1.644  &2.3$\pm$0.4    &  $\cdots$  \\
\hline\\
\end{tabular}
\\
Notes: $^{a}$S/N between 2 and 3.\\
\end{table*}


\end{document}